\documentclass[twocolumn]{aastex63}
\usepackage{amsmath}
\usepackage{amssymb}
\usepackage{graphicx}
\usepackage{color}
\usepackage{float}

\definecolor{darkgreen}{rgb}{0,0.35,0}
\shorttitle{Migration of Gas Giants}
\shortauthors{Chen et al.}
\begin{document}

\title{Retention of Long-Period Gas Giant Planets: Type II Migration Revisited}
\author{Yi-Xian Chen}
\affiliation{Department of Physics, Tsinghua University, Beijing, 100086 China} 

\author{Xiaojia Zhang}
\affiliation{Planetary Environment and Asteroid Resource Laboratory, Origin Space Co. Ltd., China}

\affiliation{Department of Astronomy \& Astrophysics, University of California, 
Santa Cruz, CA 95064, USA}

\author{Ya-Ping Li}

\affiliation{Theoretical Division, Los Alamos National Laboratory, Los Alamos, NM 87545, USA}

\author{Hui Li}

\affiliation{Theoretical Division, Los Alamos National Laboratory, Los Alamos, NM 87545, USA}

\author{Douglas. N. C. Lin}

\affiliation{Department of Astronomy \& Astrophysics, University of California, 
Santa Cruz, CA 95064, USA}

\affiliation{Institute for Advanced Studies, Tsinghua University, Beijing 100086, China}

%\email{yx-chen17@mails.tsinghua.edu.cn}
%\date{\today}

\begin{abstract}
During their formation, emerging protoplanets tidally interact with their
natal disks. Proto-gas-giant planets, 
with Hills radius larger than the disk thickness, open gaps and quench 
gas flow in the vicinity of their orbits. It is usually assumed that 
their type II migration is coupled to the viscous evolution of the disk. 
Although this hypothesis provides an explanation for the origin of 
close-in planets, it also encounter predicament on the retention of
long-period orbits for most gas giant planets. Moreover, numerical
simulations indicate that planets’ migrations are not solely 
determined by the viscous diffusion of their natal disk. Here we
carry out a series of hydrodynamic simulations combined 
with analytic studies to examine the transition between different paradigms of type II 
migration. We find a range 
of planetary mass for which gas continues to flow through a severely 
depleted gap so that the surface density distribution in the disk region 
beyond the gap is maintained in a quasi-steady state. The associated gap profile modifies the location of corotation \& Lindblad 
resonances. In the proximity of the planet's orbit, high-order Lindblad \& corotation torque are weakened by the gas depletion in the gap while 
low-order Lindblad torques near the gap
walls preserves their magnitude. Consequently, the intrinsic surface 
density distribution of the disk determines delicately both pace and direction of planets' type II migration. We show that this effect might stall the inward migration
of giant planets and preserve them in disk regions where the surface density is steep.
%The combined effect of diffusive flow across the gap,
%narrow gap structure, and the Lindblad torque balances sustain
%a quasi steady state and retain the gas giants near the site of their formation.
\end{abstract}
\keywords{protoplanetary disks, planet-disk interactions, planet migration}

\section{Introduction}

The discovery of close-in gas giants, commonly referred to 
as hot Jupiters \citep{Mayor_Queloz_1995}, rekindle the theoretical
hypothesis that planets may migrate extensively during their
formation in their natal disks \citep{GT1980, Lin_Papaloizou_1986,Linetal_1996}. 
This process is driven by the planets' excitation 
of density waves near their Lindblad and corotation resonances
\citep{Goldreich_Tremaine_1982}.  The waves interior/exterior to
the planets' orbits carry a negative/positive angular momentum 
flux. The dissipation of these waves leads to a net torque
\citep[][for a review]{Papaloizou_Lin_1984,Lin_Papaloizou_1993, 
KleyNelson2012}. 

At the location where the waves are dissipated, the 
deposition of angular momentum flux into the disk gas 
modifies its local surface density ($\Sigma$) distribution 
\citep{Takeuchi1996, RafikovGoodman2001}. Although tidal perturbation
of planets with masses a few times that of the Earth
(commonly referred to as super Earths) on the structure
of typical protostellar disks is insignificant, the 
unperturbed $\Sigma$ distribution generally
introduces a differential torque. Imbalance between 
the torque exterted on the disk regions interior and
exterior to the planets' orbit induces either gain 
or loss of their angular momentum and consequently 
type I orbital migration \citep{Ward_1986, Ward_1997}. 
This process may be relevant to the 
origin of super Earths with periods 
ranging from days to weeks \citep{Terquem_Papaloizou_2007}.
Through an extensive series of numerical simulations, 
\citet{Paardekooperetal2010a,Paardekooperetal2011} 
determined the magnitude of nonlinear Lindblad 
($\Gamma_L$) and corotation ($\Gamma_C$) torque associated with 
the density waves excited near the Lindblad resonances and 
co-orbital region respectively. They also constructed a set of 
useful approximation for $\Gamma_L$ and $\Gamma_C$ as 
functions of the planets' mass ratio with their host stars $q$
and semi major axis $r$, disk's aspect ratio $h$, surface density
$\Sigma$ and mid-plane temperature $T$ distribution and effective
viscosity $\nu$ \citep{ShakuraSunyaev1973, Kretke_Lin_2012}.

%The discovery of a close-in hot Jupiter, 51 Peg b\citep{Mayor_Queloz_1994} rekindles the hypothesis that some gas giants may have migrated extensively to the proximity of their host stars \citep{Linetal1996}.  

For giant planets with masses comparable to that of Jupiter, they
exert strong torques in disk regions near their orbits which lead
to the type II migration \citep{Lin_Papaloizou1986a}. 
In classical type II migration, a deep gap is formed when 
the planets' mass exceeds a thermal limit with which their
Hills radius becomes larger than the disk scale height.  This
gap not only severely depletes gas in the planets' co-orbital region 
but also effectively quenches flow across the gap \citep{Brydenetal2000}.  
In this limit, planets interact with the disk solely through the
Lindblad torque and their evolution is locked to the viscous 
diffusion of the disk gas. Interior/exterior 
to the radius of maximum viscous stress ($r_{cr}$) 
\citep{LydenBellPringle1974}, planets undergo orbital decay/expansion 
respectively with the viscous diffusion of the disk gas. Provided 
the characteristic mass of the unperturbed region of the disk
exceeds $M_p$, the embedded planet may undergo substantial 
migration on the disk's evolution timescale \citep{Lin_Papaloizou_1986}.

Although type II migration 
is generally slower than that of the type I migration, it still
poses a challenge to the retention of a majority of gas giants as
cold Jupiters in line with gas giants' observed period distribution.
\citet{Nelsonetal2000} showed that a Jupiter formed at 5AU in 
a viscously evolving disk may undergo type II migration on a 
$\sim 10^5$ yr time scale. Population synthesis models \citep[e.g.][]{Ida2013}
require a ten-fold reduction in the type II migration rate to match gas
giant planets' period distribution.  
In principle, gas giants formed outside $r_{cr}$ 
may initially migrate outwards and be retained. But, the apparent 
sizes of typical protostellar disks \citep{Andrewsetal2016} are 
generally much larger than the orbital semi major axis of most 
known cold Jupiters.  In these disks, $r_{cr}$ are likely to have expanded
beyond the semi major axis of the gas giants, which would 
lead to their orbital decay.  In the period distribution of 
cold Jupiters, there is no indication of significant excess of
distant cold Jupiter \citep{Vignaetal2012}.  These observational 
properties do not support the scenario that most cold
Jupiters may be retained due to an outward orbital migration beyond $r_{cr}$.

However, recent hydrodynamical simulations show that the gap opened by planets exceeding the 
thermal mass is not totally depleted. The diffusion of residual materials maintains a steady 
gas profile \citep{Duffel_etal_2014,Fungetal2014,DurmannKley2015,Robert2018}. As a result, gap-opening 
planets might not be coupled with the viscous viscosity predicted by classical type II theory. 

Using semi-local deposition descriptions of planetary torque (assuming main contributions from within the gap), \citet{Kanagawaetal2015MNRAS} and \citet{Duffell2015b} constructed self-consistent description of gas profiles from first principle that fits with simulations. Based on such scalings, \citet{Kanagawaetal2018} provided an empirical formula for the migration rate which extends type I migration torque scalings \citep{Paardekooperetal2010a} to the type II region. \citet{Ida2018} 
suggested that \citet{Kanagawaetal2018}'s paradigm can slow down migration if the viscosity parameter for gap-opening is much smaller than the global $\alpha$ for accretion. However, outlier cases still exist in which the torque deviates markedly from the inferred value.  Certain simulated models also produce outward migration, 
which are consistent with some previous simulations \citep{Massetetal2006,Duffell2015}. 

It was suggested that such suppression or reversal of inward migration might be the effect of non-linear corotation torque. %However, when the surface density within the horseshoe region is reduced, corotation torques, even for optimum viscosity, should generally become sub-dominant \citep{Dempseyetal2020} or at least scale as the Lindblad torque. 
Here we propose that for gap-opening planets, an explanation of such phenomenons may arise from the excitation and balancing of Lindblad resonances instead. Non-linear scatter in corotation torque, even for optimum viscosity case where vortensity gradient is preserved, usually does not exceed the order of linear corotation torque itself \citep{Masset2001,Paardekooperetal2011}. Since the corotation radius exists within the flat bottom of the gap, corotation torque should scale with $\Sigma$ together with the Lindblad torques \textit{within} the gap. They 
are naturally dominated by the low-order Lindblad torques risen from \textit{outside} the gap, if any, where the surface density is higher by 1-2 orders of magnitude, as suggested by analytic gap modeling with more improved non-local deposition descriptions \citep{GinzburgSari2018,Dempseyetal2020}. Due to the very reason that these ``rogue" resonances are hard to trace with {\it ad hoc} descriptions, even the most recent analytical models suffer from inaccuracies for deep gaps opened by planets exceeding thermal mass.
%thermal mass planets
%Nevertheless,  \citet{GinzburgSari2018}'s analysis does not apply to thermal mass planets in disks with high viscosity. \citet{Dempseyetal2020} revised the description to include a pileup of gas outside the planet's orbit, and found that this pileup always accompanies the inward migration of the planet,  and potential outward migrations are left unaccounted for. In fact, according to their formula, the pileup should naturally disappear or appear on the inside for outward migration, consistently giving more potentials for inner torque to cancel out outer torque - such cases are not discussed. 

In this work, we use 2D \texttt{FARGO}, 2D and 3D \texttt{LA-COMPASS} simulations  %When torques turn out to be slightly positive or negative, pileups should not occur self-consistently even if we impose \citet{Dempseyetal2020}'s term in boundary condition. 
to show that with a relatively steep gradient in the $\Sigma$ distribution, type II migration of a Jupiter-mass planet may be significantly suppressed and even reversed. We also trace how materials efficiently diffuse through the gap by loading on to horseshoe streamlines and librate across the horseshoe region, and develop a detailed model for maintenance of gas flow across the gap. 

In such cases, the actual migration direction relies delicately on the balance between the low-order Lindblad torques and the surface density at the resonances.  Our results account for the outward migration under certain circumstances and the deviations of simulated migration rate from \citet{Kanagawaetal2018}'s formula. We avoid uncertainties introduced by first-principle analytic approximation of the $\Sigma$ profile around the planet, by computing the Lindblad and corotation torques with simulation-generated $\Sigma$ distribution. Each $m$-th order resonance's location and its torque contribution are calculated discretely based on \citet{1993ApJ...419..155A} and \citet{Ward1989}'s formulae that provide very close fit to numerical calculations \citep{KorycanskyPollack1993}. We emphasize that the $\Sigma$ slope in the weakly perturbed regions 
of the disk (beyond the gap's edges) plays a large role in the balancing of lower-order torques when they dominant the total torque.

The paper is organized as thus: in \S \ref{sec:methods} we introduce the methods of study, including the hydrodynamical simulation setup and the calculation of resonance torques from steady gas profiles. In \S \ref{sec:results} we present results we obtained from our main 2D simulations with \texttt{FARGO} to show how changes in the gas profile induce fluctuations in low order Lindblad torque and delicately control the migration direction. In \S\ref{sec:LA-COMPASS} we present results from some additional 2D and 3D simulations with \texttt{LA-COMPASS}.  In \S\ref{sec:evolution}, we consider the long term mass and orbital evolution of emerging 
gas giant planets and in \S \ref{sec:summary} we summarize and discuss some outstanding issues.

\section{Methods}
\label{sec:methods}

\subsection{Hydrodymamical Model}
In our simulations, we use a geometrically thin and non-self-gravitating proto-planetary disk (PPD). We choose a 2D cylindrical coordinate system $(r,\varphi)$, and the origin locates at the position of the $M_*=M_{\odot}$ central star. We adopt a simple model in which the PPD’s temperature is independent of the distance above the midplane $T_{disk}\propto r^{-l}$ with aspect ratio

\begin{equation}
    h=\frac{c_{s}}{v_{K}}(r)=\frac{H(r)}{r}=h_{p}\left(\frac{r}{r_{p}}\right)^{(1-l)/2}
    \label{eq:hr}
\end{equation}
where {$v_K$ is the Keplerian velocity}, $H$ is the scale height, $c_s$ is the sound speed. $r_p$ is the planet's radius of circular orbit and $H_p=r_p h_p$ is the scale height at $r_p$. The disk has an initial profile of 
\begin{equation}
    \Sigma(r)=\Sigma_{p}\left(\frac{r}{r_{p}}\right)^{-s}
    \label{eq:sigmar}
\end{equation}
where $\Sigma_p$ is the surface density at the orbital radius of the planet, $l$ and $s$ are power indices. 
The 2D velocity vector of the gas is $\vec{v}=\left(v_{{r}}, {v}_{\varphi}\right)$, and angular velocity is $\Omega=v_{\varphi}/r$.

By numerically solving the continuity equation
\begin{equation}
   \frac{\partial \Sigma}{\partial t}+\nabla \cdot(\Sigma \vec{v})=0,
   \label{eq:continuity}
\end{equation}
and the equation of motion
\begin{equation}
    \frac{\partial \vec{v}}{\partial t}+\vec{v} \cdot \nabla \vec{v}=\frac{\nabla \mathcal{P}}{\Sigma}-\nabla \phi+\vec{f}_{\nu},
    \label{eq:equationofmotion}
\end{equation}
where $\vec{f}_{\nu} $ represents the viscous force per unit mass associated with stress tensor \citep[e.g.][]{Nelsonetal2000} and $\mathcal{P}$ is the vertically integrated pressure.

The total potential is given by \citep[e.g.][]{GT1980}
\begin{equation}
    \phi=\phi_p-\frac{G M_{\mathrm{*}}}{\left|\vec{r}\right|}; \ \
    \phi_p=q \Omega^{2}(r) \vec r_{p} \cdot \vec r -\frac{G M_{\mathrm{p}}}{(\left|\vec{r}_{{p}}-\vec{r}\right|^2+\epsilon^2)^{1/2}}
    \label{potential}
\end{equation}
where $q \equiv M_{{p}}/M_{\mathrm{*}}$ is the planet-to-star mass ratio, 
$\epsilon=0.4R_H$ is the softening length we adopt, and $R_H =(q/3)^{1/3} r_p$ is the planet's Hill radius.

We first use \texttt{FARGO} code \citep{Masset2000} to integrate the dynamical equations, covering a range of [$r_{min}$, $r_{max}$] in radial direction and [$0$, $2\pi$] in azimuthal direction. In a cylindrical coordination, the disk is logarithmically divided 
into 1024 grids in radial and averagely divided into 1256 grids in azimuthal which is considered as our standard resolution in this paper. 
We use the damping boundary condition \citep{deValborroetal2006} to provide wave killing zones at each radial edge of the disk to prevent wave reflections. %The disk is disturbed by the planet's potential softened with a length of $\epsilon=0.6H$, where $H$ is the local disk height. 
In the first 600 orbits the disk was disturbed by the fixed-orbit planet. 
After that the planet was released to interact with the disk.  

We adopt the conventional $\alpha$ prescription of the kinematic viscosity \citep{ShakuraSunyaev1973}
\begin{equation}
    \nu=\alpha_{\nu} c_s H = \alpha_\nu h^2 \Omega r^2
\end{equation}
for the disk region where $r$ is much smaller than the characteristic disk size. 
In most of the models, the viscosity parameter $\alpha_\nu$ is assumed to be a 
constant of radius.  In one particular model, we also consider the possibility
that $\alpha_\nu \propto r$ (see \S\ref{sec:2dlacom}).  The radial velocity and disk accretion rate 
due to viscous diffusion are given as \citep[see][for a review]{Franketal1992}:
 
\begin{equation}
v_r \sim  \frac{3 \nu}{2 r} = \frac{3 \alpha_\nu h^2 \Omega r}{ 2}, 
\label{eq:vrvis}
\end{equation}

\begin{equation}
\dot M \sim 3\pi \nu \Sigma = 3\pi \alpha_{\nu}h_p^2 r_p^2\Sigma_p \Omega_p \cdot 
\left(\frac{r}{r_p} \right)^{1.5-s-l}
\label{eqn:accretion rate}
\end{equation}
for a constant $\alpha_\nu$.

In \S\ref{sec:results}, we adopt $s+l=1.5$ for initial conditions such that $\dot M$ is independent of $r$ which gives a steady 
accretion state of the disk. 
%In our main runs, we fix $h_p=0.03, \alpha=1.1\times 10^{-2}$ such that $\nu_{p}=10^{-5}\left[r_{p}^{2} \Omega_{p}\right]$, 
The normalization factors for ${\dot M}$, $\nu$, and $\Sigma$ can be expressed, respectively, in code units 
($M_\ast$, $r_p$, and $\Omega_p={\sqrt{GM_\ast/r_p^3}}$) such that ${\dot M}= 9.4\times 10^{-8} m_* \Omega_{p} \nu_{-5}\Sigma_{0,-3} M_{\odot} {\rm yr}^{-1}$ where $m_*=M_*/M_{\odot}$,
$\nu_{-5}=\nu(r_p)/(10^{-5}r_{p}^{2} \Omega_{p})$ and $\Sigma_{0,-3}=10^{-3}M_*/r_p^2$. The dimensionless viscosity would be $\alpha_\nu = 10^{-5} \nu_{-5} / h_p^2$. We trace the evolution of a Jovian planet ($q=10^{-3}$ for $M_*=m_\odot$) in disks with different initial steady-accretion surface density gradients and parameters ${\dot M}$, $\Sigma_p$, and $h_p$. 
The values of these parameters are chosen to be in the range expected for typical PPDs and they are listed in Table \ref{models}.% with Model 1 
%as a reference run for detailed analysis.

\begin{table*}
	\centering
	\caption{Simulation parameters used in this paper.}
	\label{models}
	\begin{tabular}{cccccccc} % four columns, alignment for each
		\hline
		\hline
		Model & $r_p$[AU]  & ${\dot M}[M_{\odot}/\mathrm{yr}]$ & $h_p$ & $\alpha_\nu$ & $s$ & $r_{min}[r_p]$ & $r_{max}[r_p]$\\
		\hline
     1   &1 & $6 \times 10^{-7}$  &  0.03 & $1.11\times 10^{-2}$ & 0.5, 1, 1.5, 2 & 0.25 & 8.0\\
      2&  5.2 & $1 \times 10^{-7}$   & 0.05  &  $3\times 10^{-3}$   & {0.5, 1} & 0.25 &8.0 \\
      %3&   $10^{-7}$  & 0.03  &    $8.33\times 10^{-3}$   &  0.5, 1, 1.5\\
        \hline
	\end{tabular}
\end{table*}

\subsection{Reduction of Circumplanetary Torque}
\label{reduction}
The initial power-law $\Sigma$ distribution is appropriate for a steady state disk in the
absence of any embedded planets.  Our simulations begin with a disk adjustment stage
in which the disk evolves under the influence of a non migrating planet.  This relaxation
in disk structure ensures that we do not introduce an impulsive perturbation which can lead to an artificial
type III migration \citep{PapaloizouMasset2003}.  Just before the planet is released, a deep gap is 
carved around the planet.  The minimum surface density in the gap $\Sigma_{min}$ is 
about 2-3 orders lower than that of the unperturbed disk. 
This gap region corresponds to the corotation region with the half-width roughly equals to 
$R_H$. An accumulation of the disk materials near the planet's vicinity also gives rise 
to a torque contribution $\Gamma_D$ from the circumplanetary disk (CPD).

The magnitude of $\Gamma_D$ depends on the amount of disk materials which interacts with the planet. 
Due to its close proximity to the planet, this part of torque can lead to unstable migration 
since a small amount of asymmetry in the CPD can result in a strong variation in the torque. But, if the 
streamlines in the CPD are closed inside the planet's Hill's radius, the CPD's time-average torque on 
the planet would vanish in the asymptotic limit. 

\citet{DAngeloetal2005} carried out numerical simulations to investigate the impact of the torque arising within the Hill's radius $R_H$
on type II migration rate. They indicated that any inaccuracy in calculation of this torque can introduce spurious torques. In order to avoid this problem, we neglect the disk gas accretion onto the planet and exclude the torque from the CPD similar to previous studies \citep{Fungetal2014,Kanagawaetal2018}. In our studies, we specifically exclude planetary torque from the region within $0.6R_H$ of the planet.
%Therefore, due to our limited simulation resolution, in order to avoid this we reduce the torque by removing $10\%$ of the disk mass within $R_H$ in each time step 
%({\color{red} how big is each timestep?}) during the migration. After about 100 orbits the surface density in the vicinity of the planet can be significantly reduced and be steady. Instead, one can also achieve this by turning on the planet's accretion. Here we just remove the mass from the disk and do not add the mass and angular momentum onto the planet. 

\subsection{Obtaining Torque Components From Gas Profiles}
\label{torqueanalysis}
From an established $\Sigma$ distribution \textit{directly} obtained from numerical simulations, we calculate
the torque contribution from each Lindblad and corotation resonance assuming waves are linear, following the
analytical formulae from \citet{GT1980,Ward1989,1993ApJ...419..155A}.  Previous treatments often separated the 
Lindblad and corotation torques depending on whether it arises from inside or outside the corotation/horseshoe 
region \citep[e.g.][]{PaardekooperPapaloizou2009,2018MNRAS.473.5267M}. However, such linear approximation may not be valid for Jovian 
planets with a mass comparable to the thermal limit. For these planets, the co-orbital region has a half-width of
$\sim H$ and it is comparable to the gap half-width, while high-order Lindblad resonances converge at distance 
within $\sim 2H/3$ from the planet (see below subsection). Therefore we should not expect these torques to have a clear boundary in
their spatial distribution. \citet{Dempseyetal2020} proposed that one may be able to separate them by looking at 
the distribution of \textit{deposited torque} instead of excited torques. 

The differences between the excited torque and deposited torque have been elaborated by \citet{Takeuchi1996,  RafikovGoodman2001,2002ApJ...569..997R}. The waves excited by the planet at resonant locations propagate over a 
distance before they dissipate and deposit angular momentum in the disk. Whereas wave excitation controls the 
planet's migration, it is the deposition that directly moulds the surface density profile of the disk. The propagating flux of angular momentum carried the density waves and their non local dissipation in 2D \citep{Papaloizou_Lin_1984, Lin_Papaloizou1986a} or 3D \citep{Lin1990} limits have been studied with numerical simulations, but it's still very useful to construct steady profiles from first principles, by equating the viscous torque
and the deposited torque on any infinitesimal section of surface profile $\mathrm{d}j (\Sigma(r)) =\mathrm{d} T_{dep}(\Sigma(r))$. 

Under assumption that most of deposited torques are released at the bottom of the gap with effective bottom density $\mathrm{d}T_{dep}(\Sigma(r))\approx\mathrm{d}T_{dep}(\Sigma_{min})$, one obtains a similar expression of $\Sigma_{min}$ compared to \citet{Kanagawaetal2015MNRAS}, who approximated deposition torques with excited torques to obtain
\begin{equation}
    \frac{\Sigma_{min}}{\Sigma_{p}}\approx\frac{1}{1+0.04 K} \ \ \ \ \ \ {\rm where}
    \ \ \ \ \ \ K \equiv q^2h_p^{-5}\alpha^{-1}.
    \label{bottomdensity}
\end{equation}
This approximation has been shown to fit well empirically with simulation data of steady gap profiles, especially non-accreting planets which open partial gaps \citep{DM13,Duffell2015,Kanagawaetal2015,Duffell2020,Dempseyetal2020} where dominating torques self-consistently ``live" within the gap. 

\citet{Kanagawaetal2018} applied Eqn \ref{bottomdensity} to migration, proposing that the migration of a gap-opening planet should, analogous to Type I migration reference torque $\Gamma_0=\left(q^{2} / h_p^2 \right) \Sigma_{p} \Omega_{p}^{2} r_{p}^{4}$ \citep[e.g.][]{Tanakaetal2002}, scale as
\begin{equation}
\Gamma_{ref}=\left(\frac{q}{h_p }\right)^2 \Sigma_{min} \Omega_{p}^{2} r_{p}^{4}=\Gamma_0\frac{\Sigma_{min}}{\Sigma_{p}}.
\label{ref}
\end{equation}

This prescription naturally bridges the gap between type I and type II migration, and it accounts for migration speeds obtained from many numerical simulations. However, in certain cases the torque obtained from numerical simulations still deviates considerably from the above expected values. In some cases, simulations show outward migration \citet{Massetetal2006,Duffell2015b} for the gap-opening planets. Based on their identification of asymmetry in the torque distribuion throughout the horseshoe flow (see their Fig. 10), \citet{Kanagawaetal2018} suggested that these disparities might be the effect of non-linear corotation torque. However, as we have indicted above, it may not be appropriate to measure the effect of corotation torque and differential Lindblad torque under the assumption of spatial separation. Usually, non-linear scatter in the corotation torque, even for optimum viscosity case where vortensity gradient in the unperturbed flow is preserved, does not exceed the order of linear corotation torque itself \citep{Masset2001,Paardekooperetal2011}.  Since corotation radius always exists within the flat bottom of the gap, the upper limit of corotation torque should scale together with the Lindblad torques to be $\propto\Sigma_{min}$. 

Here we show that for our gap-opening cases, it is the low-order Lindblad torques at or just beyond the edge of
the gap that is controlling the migration speed and direction of the planet. These torques naturally dominate if 
they arise from regions where the surface density is not appreciably perturbed, {\it i.e.} 1-2 orders of 
magnitude larger than $\Sigma_{min}$.  This $\Sigma$ profile itself deviates from the prediction of Eqn 
\ref{bottomdensity} since the underlying assumption $\mathrm{d}T_{dep} (\Sigma(r)) \approx 
\mathrm{d}T_{dep}(\Sigma_{min})$ breaks down. 

Direct evaluation of the torque densities from the $\Sigma$ distribution across the gap requires non-trivial 
separation of Lindblad torques and corotation torques.  First-principle analytic models of planet gaps are 
incomplete for very deep gaps \citep{GinzburgSari2018,Dempseyetal2020} due to assumptions that go into torque 
deposition theories. Here we adopt an alternative approach by taking established $\Sigma$ and corresponding 
$\Omega$ profiles from simulations and use them to calculate the corresponding excited torques. This method 
bypasses the uncertainties in the deposited torques that mould the $\Sigma$ profiles. It is straight forward to 
separately compute the location and contribution of each $m$-th order corotation and Lindblad torque. The radial 
profiles for the torque calculations are azimuthally averaged from a 2D distribution excluding the CPD region within $0.6R_H$, to
be consistent with the exclusion of $\Gamma_D$ in \S \ref{reduction}.

\subsection{Lindblad torque}

In a disk with radial surface density profile $\Sigma(r)$, and corresponding angular velocity profile $\Omega(r)$, the Lindblad resonances are located at $r_m$ where 

\begin{equation}
    m[{\Omega(r)-\Omega_p}]=\pm \kappa \sqrt{1+m^2h(r)^2} \equiv \pm\kappa',
\end{equation}
with the plus/minus sign for inner/outer Lindblad resonance (ILR/ORL). The epicycle frequency is $\kappa=\sqrt{4\Omega B}$ and the Oort constants $A=\dfrac r2\dfrac{\mathrm{d}\Omega}{\mathrm{d} r}$
and $B=\Omega+A$.  We define $\Delta_m=(r_p-r_m)/r_p$. 
The $m$-th order Lindblad torque is given by:

\begin{equation}
    \Gamma_m= \left.\dfrac{m{\pi^2}\Sigma}{r\mathrm{d}D/\mathrm{d}r}
    \left[r\dfrac{\mathrm{d}\phi_m}{\mathrm{d}r}+2{m^2}(1-\dfrac{\Omega_p}{\Omega})\phi_m
    \right]^2f_L\right|_{r_m}.
    \label{torque}
\end{equation}
where $D=\kappa^2-m^2(\Omega-\Omega_p)^2$, and 

\begin{equation}
    f_L=\left[\sqrt{1+m^2h^2}(1+4m^2h^2) \right]^{-1}
\end{equation}
is a cutoff factor due to the shift in resonances induced by pressure \citep{1993ApJ...419..155A,KleyNelson2012}. We have expanded the smooth potential of the perturber (Eqn \ref{potential}) as

\begin{equation}
    \phi_p=\sum_{m=0}^{\infty} \phi_{m}(r) \cos \left\{m\left[\varphi-\varphi_{\mathrm{p}}\right]\right\}.
\end{equation}

We use the exact form of $\phi_m$ and $\dfrac{\mathrm{d}\phi_m}{\mathrm{d}r}$ instead of applying analytical approximations for small $|\Delta_m|$ \citep{GT1980}. Since the bulk of torques may come from low-order $m$ resonances in the deep gap case, evaluation of the discrete torques can minimize inaccuracies which often accompany the analytical expression of \textit{torque densities}. 

%Although our torque analysis is mainly centered around 2D scenario, Eqn \ref{torque} can be obtained from a 3D analysis with some approximations have been made in integrating across the vertical direction. Such approximations are valid for smooth perturbing potential such as adopted in our context \citep{1993ApJ...419..155A}.

\subsection{Corotational Torque}
The corotation radius $r_c$ is given by the requirement

\begin{equation}
 {\Omega(r)-\Omega_p}=0,
\end{equation}
In unperturbed disks with power law density profile $\Sigma(r)\propto r^{-s}$, there is a shift of $r_c$ from $r_p$ by an amount $\Delta_c \equiv (r_p-r_c)/{r_p}\approx s h_p^2/3$. 

The $m$-th component of the linear corotation torque is given by

\begin{equation}
    \Gamma_{m}^{\mathrm{C}}=\left.\frac{m \pi^{2}}{2} \frac{\phi_{m}}{r d \Omega / d r} \frac{d}{d r}\left(\frac{\Sigma}{B}\right)f_C\right|_{r_c},
    \label{eq:gammaco}
\end{equation}
where $\Sigma/B$ is the vortensity, $f_C(r)$ is the cutoff factor of corotation torques given by \citet{Ward1989}: 

\begin{equation}
    f_C=\frac{1+\xi^{2}\left(A^{2} / \Omega B+4.8\right)}{\left[1+\xi^{2}\left(A^{2} / \Omega B+1-2 A / B\right)\right]^{2}}, \ \ \xi=\dfrac{mh\Omega}{\kappa},
\end{equation}
and all components are evaluated at the corotation radius for a planet following circular orbit. 

The cutoff functions that we apply to the Lindblad torque and corotation torque are verified by \citet{KorycanskyPollack1993} to fit very well for the ($m<1/h$) dominant resonance torques 
numerically calculated from linear equations, and they are also consistent with \citet{Ward_1997}. 

We have verified that results of Lindblad and corotation torques in an unperturbed disk with profile $\Sigma(r)\propto r^{-s}$ is consistent with empirical estimates of type I torque \citep[e.g.][]{Paardekooperetal2010a}. The steady state torque also depends on how the corotational torques saturate with respect to the disk viscosity. In general the total torque scales with $\Gamma_0$ and the differential Lindblad torques outweighs the corotation torque in driving the planet's inward migration 
(see below). 

\section{Direction and pace of gap-opening planet's migration}
\label{sec:results}

In this section, we present the results of numerical simulation with the \texttt{FARGO} code
(see \S\ref{sec:methods}).

\subsection{Model 1: Gap profile and torque components for different disk structures}
\begin{figure}[htp!]
\centering
\includegraphics[width=0.45\textwidth]{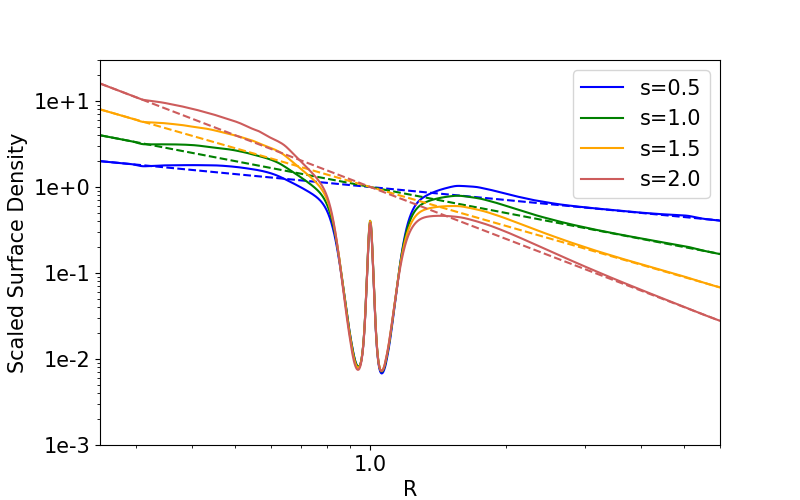} \\
\includegraphics[width=0.45\textwidth]{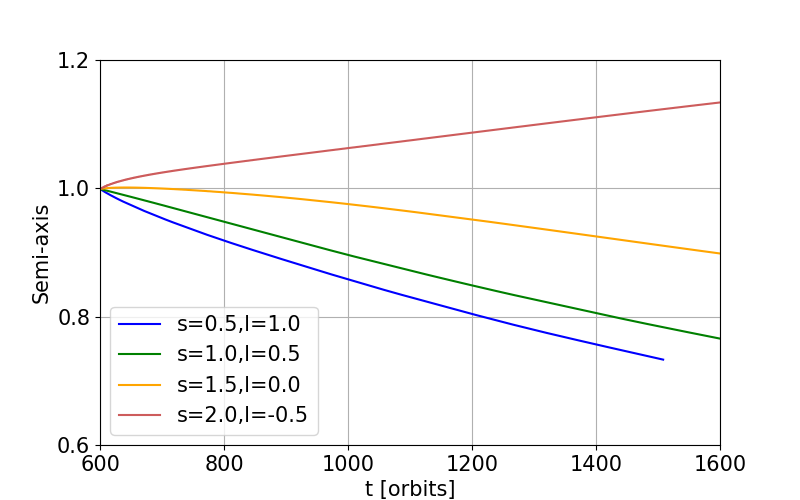}
\caption{Upper panel: {the perturbed azimuthally-averaged gas surface density (solid lines) and the orginal unperturbed profiles (dashed lines in corresponding color), measured in units of initial $\Sigma_p$ at the planet location, for the four cases of Model 1 using FARGO. The snapshots are taken at orbital time $t=600$ when we release the planet}. Lower panel: corresponding evolution of planet orbital radius after it is released, measured in units of $r_p$.}
\label{fig:model1}
\end{figure}

The simulation results of Model 1 is plotted in Fig \ref{fig:model1}. We consider four values of 
$s=0.5, 1, 1.5, 2$ for the background (i.e. initial and unperturbed) surface density slope (Eqn \ref{eq:sigmar}).
In these four constant-$\alpha$ cases, the power-law index of the temperature distribution 
(Eq \ref{eq:hr}) is taken to be $l=1.5-s$ so that, in the absence of a planet, a steady state 
is maintained. We used damping boundary conditions for both inner and outer boundaries.  In most cases, a steady
state is established such that the total disk mass reaches an asymptotic equilibrium.

In the top panels, we show the $\Sigma$ profile after
initial $t=600$ orbital periods during which the planet is held at its
initial semi major axis (\S\ref{sec:methods}).  After this epoch, the planet is released.  In the lower 
panels, we plot the evolution of its semi-major axis after the release.  

%\subsubsection{Reduced contribution from all Lindblad and corotation resonances}
\subsubsection{Migration direction and speed}
We compare the migration rates of this planet, after it has induced a relatively deep gap, with established theories. In the theory of type I migration for planets with insufficient mass to significantly perturb the $\Sigma$ profile \citep[e.g.][]{Paardekooperetal2010a}, the torques in 2D isothermal disks should scale with $\Gamma_0$, albeit the coefficient is determined by the density slope and the softening length: 

\begin{equation}
    \begin{array}{l}\dfrac{\Gamma_{L}}{\Gamma_{0}\left(r_{\mathrm{p}}\right)}=-(2.5-0.1 s+1.7 l) b^{0.71} \\ \dfrac{\Gamma_{C}}{\Gamma_{0}\left(r_{\mathrm{p}}\right)}=1.1(1.5-s) b+2.2 l b^{0.71}-1.4 l b^{1.26}\end{array}
    \label{Paardekoopertorque}
\end{equation}
where $b=0.4H_p/\epsilon=0.43$ for Model 1 and 0.73 for Model 2. For the disk parameters of Model 1, the type I torque should be $\Gamma_{tot,lin}=\Gamma_{L}+ \Gamma_{C}$ such that $C(s) \equiv - \Gamma_{tot,lin}/\Gamma_0 =(2.76-s)-(1.8-1.2s)=0.96+0.2s \sim \mathcal{O} (1)$. If we apply \citet{Kanagawaetal2018}'s extrapolation 
with $\Gamma_0\rightarrow\Gamma_{ref}$ for type II migration of
planets with sufficient mass to open gaps, the migration speed in Model 1 would be

\begin{equation}
    u_p=-C(s)\dfrac{\Gamma_{ref}}{r_p\Omega_pM_p}=-6\times 10^{-5}C(s){\Omega_p r_p \over 2\pi}.
\end{equation}

In the calculation of $\Gamma_{ref}$ from Equation (\ref{ref}), we use the bottom density from \textit{our simulations} in Figure \ref{fig:model1}. This bottom density deviates from that obtained from Eqn \ref{bottomdensity} (with $K=3384$ for Model 1) by a factor of 2. The corresponding $u_p$ implies 
that the planet would 
migrate inwards by $\sim 0.06r_p$ after a thousand orbits. If we exclude corotation torque, 
the total negative torque would be boosted and $u_p$ would become closer to the viscous evolution 
velocity in type II theory \citep[e.g.][]{Ivanovetal1999,Armitage2007} such that

\begin{equation}
    u_{p,vis}=-\dfrac{4 \pi r_{\mathrm{p}}^{2}\Sigma_p}{(4 \pi r_{\mathrm{p}}^{2}\Sigma_p+M_p)}\frac{3}{2} \frac{\nu_{{p}}}{r_{{p}}} \approx -9\times 10^{-5}{\Omega_p r_p \over 2\pi}.
    \label{u_pvis}
\end{equation}

The results of our numerical simulation for the $s=0.5$ and $s=1$ cases indicate that the planet 
migrates inward with an $u_p$ slightly (within an order unity) faster than these estimates. In the case for 
$s=1.5$, the planet undergoes outward migration initially before switching to a slow inward migration. For $s=2$, the planet undergoes outward migration. It is very informative to examine the 
actual torque components in the $s=2$ case even though its associated $l<0$ (temperature increases 
with radius) is unlikely to be physically realizable.  
If the net torque is dominated by contribution from the low order Lindblad resonances,
its magnitude and sign (initially at $r_p$) are more affected by the $\Sigma$ than the $T$ distribution
in the disk regions outside the gap.  We verify this conjecture in the $s=2$ case with a radial-dependent $\alpha_\nu$ and a positive $l$ in an extra simulation with the 2D \texttt{LA-COMPASS} code
(see \S\ref{sec:2dlacom} and Fig. \ref{fig:model1_lacompass}), where the outward direction of migration is maintained.  This situation is possible
near and beyond the outer edge of the dead zone in PPDs where the angular momentum 
transfer efficiency due to disk winds may increase with radius.  Photoevaporation
of disk may also lead to a very steep $\Sigma$ gradient. In this case, a steady gas profile may no
longer be maintained (see further discussions in \S\ref{sec:evolution} and Fig. \ref{fig:bc_lacompass}).

\begin{figure*}[htp]
\centering
\includegraphics[width=1.0\textwidth]{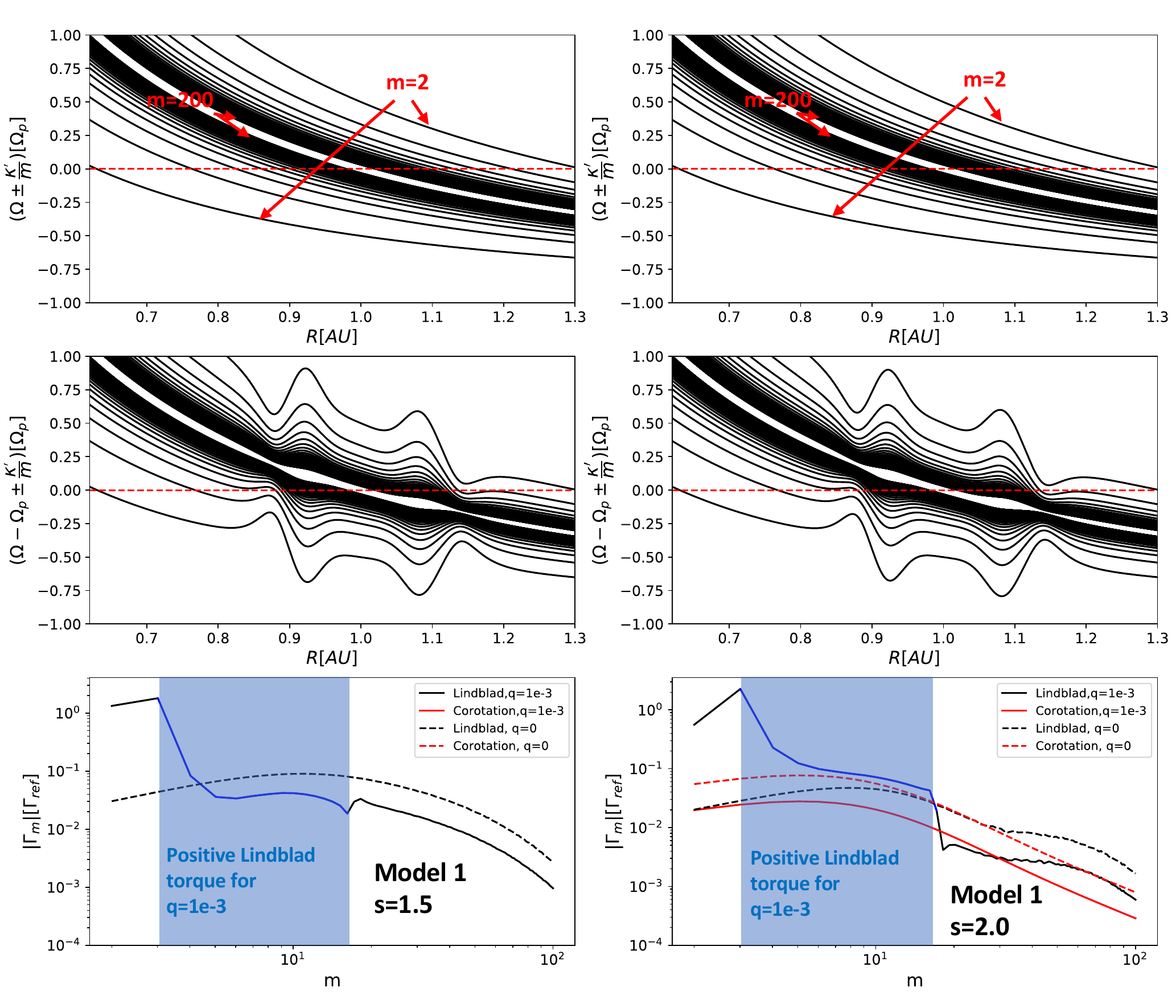}
\caption{Upper panel: the $\Omega-\Omega_p \pm\kappa'/m$ profiles for the unperturbed disk ($[\Omega_p]$ denotes the normalization unit of the axis); Middle panel: $\Omega-\Omega_p \pm\kappa'/m$ profiles for the perturbed disk with a deep gap; Lower panel: the \textit{absolute values} of each Lindblad (black lines, inner+outer torque) and corotation (red lines) torque component measured in $\Gamma_{ref}$. \textit{Outward(positive)} Lindblad torques are in blue solid lines (with highlighted background), and the components for unperturbed disk profiles are shown in dashed lines (with $\Gamma_{ref}=\Gamma_0$). Left panels are for $s=1.5$ case and right panels are for $s=2$ case in Model 1. The $s=2$ case has inward corotation torque due to the vortensity that decrease with radius.}
\label{case34}
\end{figure*}

\subsubsection{Torque distributions from numerical simulations}

The top two panels of Figure \ref{case34} show the values of $\Omega-\Omega_p \pm\kappa'/m$
(normalized with respect to $\Omega_p$) for the $s=1.5$ and $s=2$ constant-$\alpha$ cases.  
We compare the unperturbed with the perturbed profiles, showing the range of $m=2-200$. The locations of ILRs and OLRs are indicated by the null point of each profile. In the lowest panel we show the \textit{absolute value} of each Lindblad (black lines, inner+outer torque) and corotation (red lines) torque component obtained from the methods of \S \ref{torqueanalysis} for the $t=600$ profiles, measured in $\Gamma_{ref}$. In particular, the \textit{outward (positive)} Lindblad torques are plotted in blue bold lines, and the components for unperturbed disk profiles are shown in dashed lines (with $\Gamma_{ref}=\Gamma_0$ for $K=0$). The $l>0$ more realistic case of $s=2$ (\S\ref{sec:2dlacom}) has similar $\Sigma$ and $\Omega$ compared to the $l<0$ case, therefore yield similar torque distribution as the constant-$\alpha$ case since the analysis is based on established initial gap profiles.

\subsubsection{Low-order Lindblad resonances outside the gap}
Naturally, if all Lindblad resonances are located within the rather ``flat bottom" of the gap, the entire shape of the $\Gamma_m-m$ distribution would be unchanged from the unperturbed profile, only accompanied by a general suppression scaled with the reduction in surface density of the gas. This modification is more or less what happens to the unsaturated linear corotation torque: after scaling the torque with $\Gamma_{ref}=\Gamma_0\Sigma_{min}/\Sigma_p$, the entire profile $\Gamma_m-m$ does not significantly change since they lie in the bottom of the gap region, between all the OLRs and ILRs.  {For the $s=1.5$ case, the vortensity $(\Sigma/B)$ gradient in Equation (\ref{eq:gammaco}) disappears and both perturbed and unperturbed linear corotation torques are found to be 
negligible (therefore we have not shown any red lines in the lower left panel of Fig \ref{case34})}.  For the $\Sigma$ distribution with $s=2$, the corotation torques are negative in both $l = \pm 1/2$ cases.  This homologous scaling is also valid for the high-order Lindblad resonances.  

But this simple scaling is not valid for individual low-order Lindblad resonances located near or outside the gap edges. First, the asymmetry of OLR and ILR is transformed from the unperturbed case by the planet-induced gap. Second, low-order resonances tend to shift closer because epicycle frequency reduces to a minimum to ``avoid" contact with zero axis at the gap edges, before making a turn to intersect the zero axis at closer distance. For example, in Model 1 we see that this modification in $\kappa$ is just enough to shift the $m=3$ OLR into the gap but leave the $m=3$ ILR outside the gap where gas density is not significantly reduced from the unperturbed values.  The difference in $\Sigma$ makes up for the ILR's distance to the planet being larger than that of OLR by a factor of $\sim 2$ such that the ILR torque dominate over the OLR torque for $m=3$. 

For planets with insufficient mass to significantly perturb the $\Sigma$ distribution (in the context of type I
migration), the original asymmetry between inner and outer Lindblad torques arises from the fact that to second order in $1/m$, the ILRs tend to reside in regions with farther distance to the planet than the corresponding OLRs. Consequently the ILRs have smaller influences than the OLRs \citep[e.g.][]{KleyNelson2012}. But for Jupiter-mass planets surrounded by deep gaps, higher surface density of the gas at distance farther from the planets makes up for this loss. Ergo there appears to be a region (highlighted in Fig \ref{case34}) in $m$ producing positive total Lindblad torques. 

\subsubsection{Dependence on the surface density distribution}
\begin{figure}[htp!]
\centering
\includegraphics[width=0.5\textwidth]{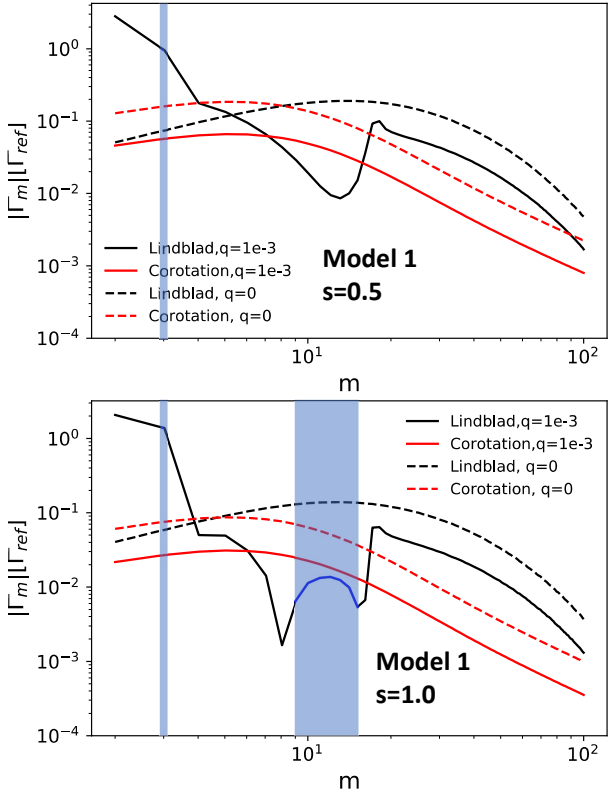}
\caption{The torque component distribution $\Gamma_m-m$ of the case in Model 1, $s=0.5$ and $s=1$.
Black and {red} lines represent the Lindblad and corotation torques respectively.}
\label{case12}
\end{figure}

For different density slope (i.e. power index $s$), the $\Omega-\Omega_p \pm\kappa'/m$ distributions as well as the asymmetries in the locations of resonances they induce are very similar (in comparison with the middle panels of Fig \ref{case34}, there are only very slight differences at the edges of the gap).  But the torque distributions are very different depending on how these asymmetries are enhanced by the $\Sigma$ profile which follows very steep distributions. For large $s$, the gas density at ILRs dominates over that at OLRs. Fig \ref{case12} shows torque components for the $s=0.5$ and $s=1$ cases in Model 1. As $s$ becomes smaller, the region of positive $\Gamma_m$ reduces (in $s=0.5$ only the very asymmetric $m=3$ resonance remains to give a positive torque), rendering a negative total torque closer to the values obtained by \citet{Kanagawaetal2018,Dempseyetal2020} through matching the observed migration rate in the simulations. Naturally, we expect the total torque to converge to \citet{Kanagawaetal2018}'s scaling when the shape of $\Gamma_m-m$ torque profile, re-normalized in $\Gamma_{ref}$, becomes similar to linear theory, since that is when most of the torques arise from the bottom of the gap. Nevertheless, since the total torque so delicately depends on the balance of the shifted ILRs and OLRs with different gas density, we still expect our azimuthally averaged 2D analysis to be subject to some uncertainties.

\subsection{Maintenance of a steady $\Sigma$ profile across the gap}
\label{maintenance}
Our torque analysis is carried out for the steady state gas profiles at $t=600$. It indicates 
that the balancing of the Lindblad torque sensitively depends on the shape of the $\Sigma$ distribution. 
In order for the disk gas to preserve this profile while the planet undergoes migration, 
there has to be sufficient gas flow across the gap. 

In our models, a steady mass flux into (out of) the disk is introduced at the outer (inner) boundary of the 
computational domain. In the absence of an embedded planet, the initial $\Sigma$ (Eq. \ref{eq:sigmar}) is 
maintained with a constant inwardly flowing mass flux.  
%\subsection{Necessary condition to maintain a steady flow across the gap}
In disks with $h_p \sim 0.03-0.05$ such as our models %(in this subsection we omit the subscript $p$ and all distance-dependent variables are evaluated at the orbital radius of the planet)
, there are residual gas 
in the gap around gas giants with 
moderate (between that of Jupiter and Saturn) masses (Fig. \ref{fig:model1}). In this case, 
gas can flow across the gap and preserve the steady-state $\Sigma$ distribution outside the gap 
region. Many previous studies have been made to confirm the existence of the mass flow \citep[e.g.][]{Fungetal2014,DurmannKley2015}, here we provide an analysis on its detailed mechanism and requirements. 

In the frame corotating with the planet, gas at the edges of the gap (at $r \sim r_p \pm R_H$) 
from the planet follows two sets of horseshoe stream lines, separated in their radial
distance from the host star by $\Delta r \sim 2 R_H$. Near the planet's azimuth, the horseshoe streamlines 
make U turns in the azimuthal direction as they swap their radial distance from the host star.
The libration timescale for these horseshoe flow around $L_4$ and $L_5$ points is  
\begin{equation}
    \tau_{lib} = {\pi  \over \Omega_p}\left({4 \over 27q} \right)^{1/2}
    ={P \over {\sqrt{27 q}}}
\end{equation}
where $P$ is the orbital period of the planet \citep{SolarSystemDynamics}. Materials on the outer horseshoe 
orbit is dynamically transported to the inner horseshoe streamline on the libration timescale. 
In the limit of negligible viscosity, they follow closed horseshoe orbits and transport back again to the outer 
orbit. However, with finite viscosity, a fraction of the disk exterior to the gap (at $r_p + R_H$) diffuses onto
the horseshoe streamlines which take them to the inner wall of the gap and where it diffuses to the disk interior
to the gap (at $r_p - R_H$) within $\sim 1$ libration timescales.  This gas element does not accumulate in and around
the gap as it is not trapped indefinitely onto the horseshoe orbits.  Across the $\Sigma$ gradient at the edge 
of the gap, the width $\delta$ of the ``track'' connecting the horseshoe streamlines is determined by the 
extent of viscous diffusion (with viscosity $\nu=\nu_p$ at $r_p$) during a libration period such that  
\begin{equation}
    {\delta} \sim {\sqrt {\nu_p \tau_{lib}}} \sim H_p (\alpha \pi)^{1/2} (4/27q)^{1/4}.
\label{eq:deltar}
\end{equation}
For Model 1 with $q=10^{-3}$ and $\alpha=1.1\times 10^{-2}$, $\tau_{lib} \sim 6 P$ and $\delta \sim 0.65 H$.

The disk materials loaded onto the outer track are directly transported to the inner track to leave the 
gap region in $\sim 1$ horseshoe liberation period, bypassing the rest of the depleted gap region.
Its effective radial velocity is
\begin{equation}
    v_{r, hs} \simeq - \Delta r/\tau_{lib} = - (3^{7/6}/\pi ) q^{5/6} \Omega_p r_p
\end{equation}
which is much faster than the viscous diffusion speed in the unperturbed regions of the disk, which is also the speed for gas diffusion across the gap:
$v_{r, \nu} = -3 \nu/2 r \simeq v_r =- 3  \alpha h^2\Omega r/2$ (Eq. \ref{eq:vrvis}).
A steady inward mass flux can be maintained with a much reduced surface density on the tracks ($\Sigma_{hr}$) as long as
\begin{equation}
    {\Sigma_{hr} \over \Sigma_p} \gtrsim {v_{r, \nu} \over v_{r, hs}} \simeq {\alpha \pi h_p^2 \over 2q^{5/6} 3^{1/6}},
\end{equation}
which is very easily satisfied since these tracks exist at the edges of the gap where gas is only partially depleted. \textit{Even if} the surface density of the track is just as depleted as the bottom of a deep as the approximated Equation \ref{bottomdensity} such that $\Sigma_{hr}\approx \Sigma_{min}\sim25\Sigma_p/K$ (mind that we have emphasized that realistic bottom density itself deviates from this scaling by order unity for Jovian planets), the steady rate of gas flow 
across the gap is maintained as long as

\begin{equation}
q \lesssim q_{ss} = (50 (3^{1/6}/\pi) h_p^3)^{6/7}= 12.5 h_p^{18/7}
\label{eq:qss}
\end{equation}
which is $q\lesssim q_{ss} = 1.5 \times 10^{-3}$ for $h_p=0.03$. 

When this criterion cannot be satisfied, the gas flow would be quenched and the planet follows the classical type II migration (See \S \ref{sec:classicaltype2}). The exact critical mass can still be larger if the horseshoe tracks are not exactly as depleted as $\Sigma_{min}$.
%This critical mass ratio $q_{ss}$ is an order of magnitude larger than the thermal mass $q=3h^3$ \citep[e.g.][]{Papaloizou1984, Ward_1997}. After the planet mass ratio exceed this value, the gas density on the horseshoe tracks might be too depleted to maintain a steady mass flow across the gap and the inner would be separated from the outer disk by the planet.

\begin{figure*}[htp!]
\centering
\includegraphics[width=0.23\textwidth]{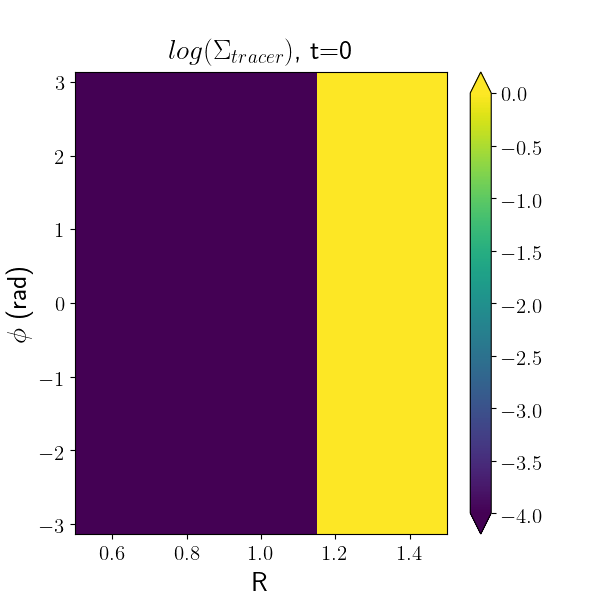}
\includegraphics[width=0.23\textwidth]{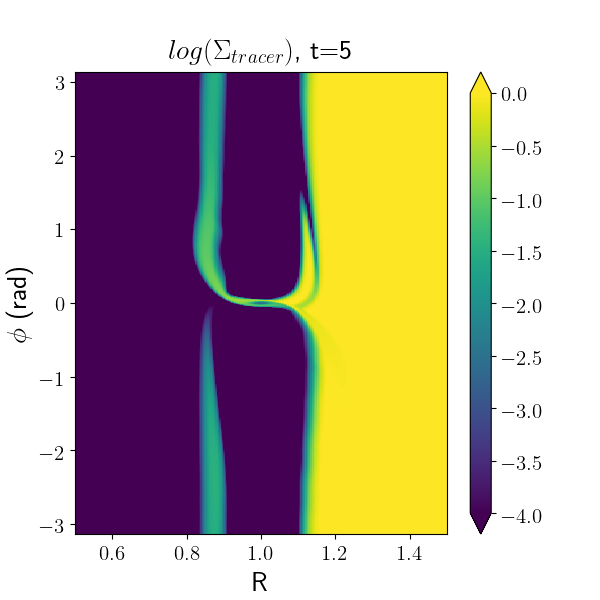}
\includegraphics[width=0.23\textwidth]{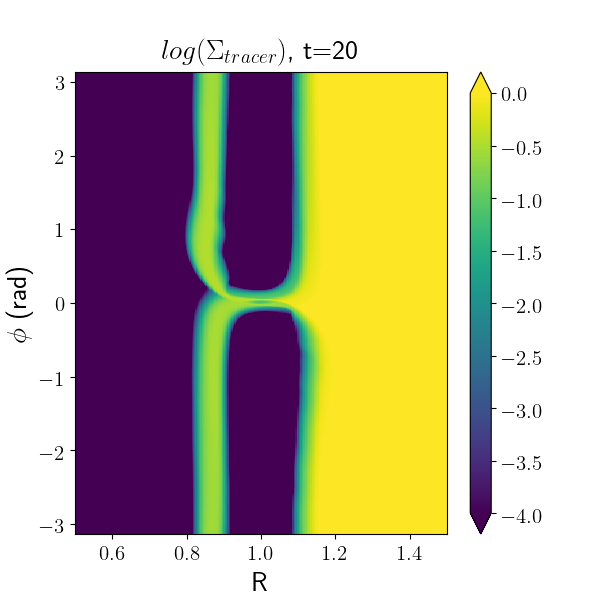}
\includegraphics[width=0.23\textwidth]{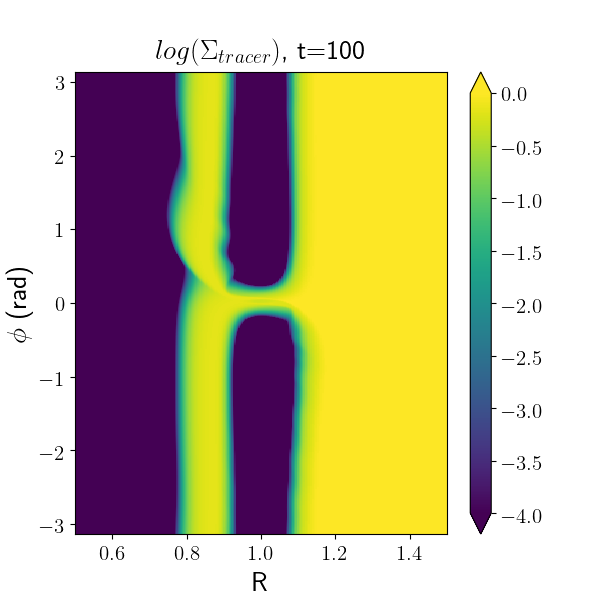}
\caption{Snapshots of tracer particle surface densities with case 3 in model1 simulated by FARGO. The tracer particles are initially located outside 1.15 from the moment at 600 orbits just before the planet is released. The panels from left to right show the tracer particle surface density at 0, 5, 20 and 100 orbits after that (t=0 corresponds to the release time of planet at 600 orbits). The planet's orbit is fixed during this simulation.}
\label{fig:tracers_fargo}
\end{figure*}

%\subsubsection{Radial extent of the connecting streamlines}
%The materials need to diffuse across the gap via contraction of their horseshoe orbit , or else they get trapped in horseshoe which gives us a typical 
Since surface density distribution of gas itself reveals little about the trajectory of materials acrossing the gap over a few libration times, we examine the details of the gas flow across the disk in Model 1 with $s= 1.5$ by uniformly distributing heuristic particles entirely outside the gap region at $t=600$, and let them completely couple with gas after planet release. We trace the surface density evolution of these particles through time to illustrate the diffusion mechanism (Fig \ref{fig:tracers_fargo}).

In the evolution of tracer particles (coupled with gas), we clearly identify connecting tracks at the edges of the entire gap region. {(Yellow) heuristic particles} are transported from the outside of the gap to the inside of the gap via these tracks, bypassing most of the inner gap regions and leaves them devoid of particles (gas). Similar diffusion patterns and typical streamlines for materials in these track structures, which distinguish them from the circulating materials (well outside the gap region) and materials in closed horseshoe orbits (trapped between these tracks) are shown in \S \ref{sec:LA-COMPASS}.

\subsection{Model 2: Case of a shallower gap}
\label{Model 2}
\begin{figure}[htbp]
\centering
\includegraphics[width=0.96\linewidth,clip=true]{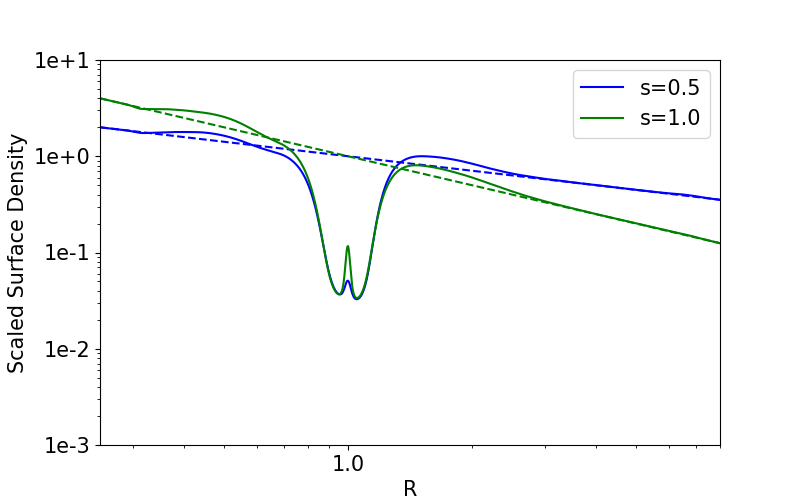} \\
\includegraphics[width=0.96\linewidth,clip=true]{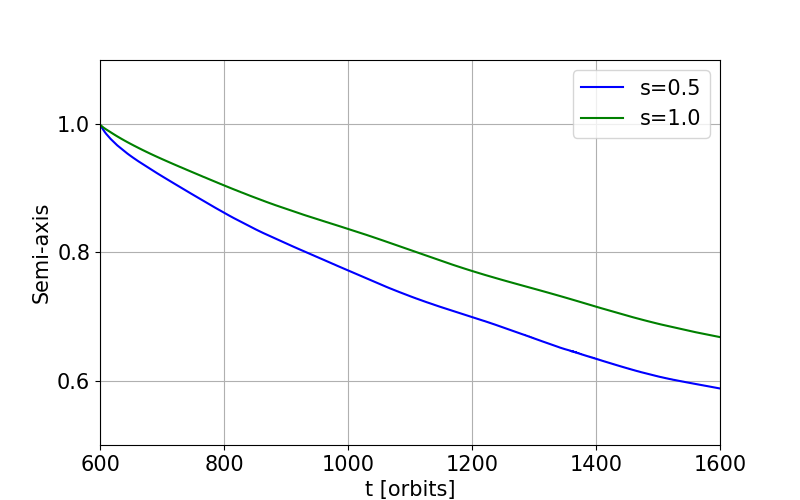}
\caption{{Top:} The surface density at $t=0$ (indicated by dashed lines) and $t=600$ orbits (indicated by solid lines)  for model 2  scaled by $\Sigma_p$, which is the initial surface density at $r_p$.
{Bottom:} The orbit evolutions for Model 2 measured in units of $r_p$, {which has $\dot M = 10^{-7} M_\cdot/yr$, $\alpha_{\nu}=0.003$ and $h_p = 0.05$ 
%with resolution of $500(n_r) \times 583(n_s)$ 
for a disk from $r_{min}=0.25r_p$ to $r_{max}=8.0r_p$. The simulations are performed with standard resolution $1024\times 1056$.}}
 \label{fig:dens_2and3}
\end{figure}

An additional series of parameters in Model 2 are simulated with $h_p=0.05$ (see Table \ref{models} and in this case $K=1067$) and a larger orbital radius $r_p=5.2$AU. {The results are plotted in Fig \ref{fig:dens_2and3}}. The values of these model parameters are 
analogous to those used by \citet{DurmannKley2015}. In this case, planet-disk interaction always 
lead to 
inward migration for a regular temperature slope, in agreement with previous results. In contrast to 
model 1 where $h_p=0.03$, the ratio $\Sigma_{\rm min}/\Sigma_p$ is larger by a factor of 3.
%(because $K \propto \alpha_\nu^{-1} h_p^{-5}$ in Eq. \ref{bottomdensity}). 
Under this circumstance, torques associated with the
high order Lindblad and corotation resonances become comparable to or larger than that due to the low-order resonance
outside the gap.  In our simulation of the $s=0.5$ case, the total negative torque in our simulation exceeds the reference torque predicted by \citet{Kanagawaetal2018} by a factor of 5.  This result is consistent with 
that presented by \citet{DurmannKley2015}. It also verifies \citet{Dempseyetal2020}'s scaling proposition 
that at least for moderate gap depth and small $s$, the realistic torques may be larger than that inferred from merely reducing the torque magnitude by a $\Sigma_p$ depletion factor (see their Fig 9). In the cases of small $s$, the realistic inward migration speed is larger than the the viscous speed and aggravates the problem of gas giant retention, only a relatively large $s$ can reduce the planetary torque.

\begin{figure}[htp]
\centering
\includegraphics[width=0.5\textwidth]{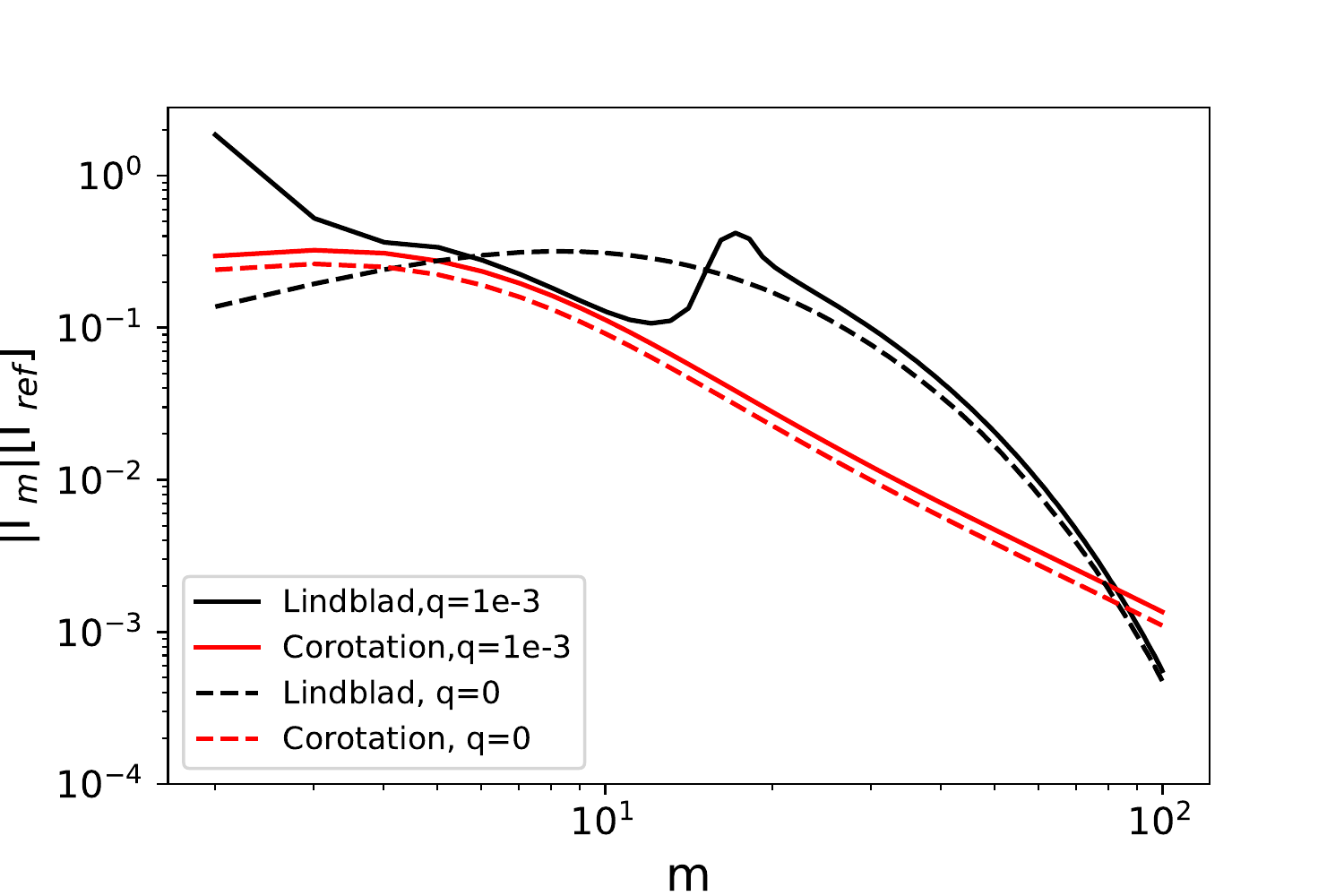}
\caption{The Lindblad (black) and corotation ({red}) torque magnitude distribution $\Gamma_m-m$ of the case in Model 2, $s=0.5$, which is close to the linear distribution scaled with $\Gamma_{ref}$ as predicted by e.g. \citet{Kanagawaetal2018}. The diagrams for $s=1$ is similar, with all Lindblad torques being negative.}
\label{fig:case1ofMODEL2}
\end{figure}

In Fig \ref{fig:case1ofMODEL2} we plot the torque components of this case, which resembles closely to the direct linear
extrapolation.  In contrast to Model 1, there is no region of $m$ that gives positive Lindblad torque in Model 2. {We deduce that for shallow gaps in low density gradient regions, the fluctuation of low-order Lindblad torques themselves is not so drastic.}

%To test for robustness, we also run Model 3 ($K=4507$) which is slightly different from Model 1 in viscosity. The results (the migration directions and torque components) are very similar to Model 1. This means that in a certain range the initial direction of migration is not chaotic to small change in the parameter space. 

\begin{figure}[htp]
\centering
\includegraphics[width=0.5\textwidth]{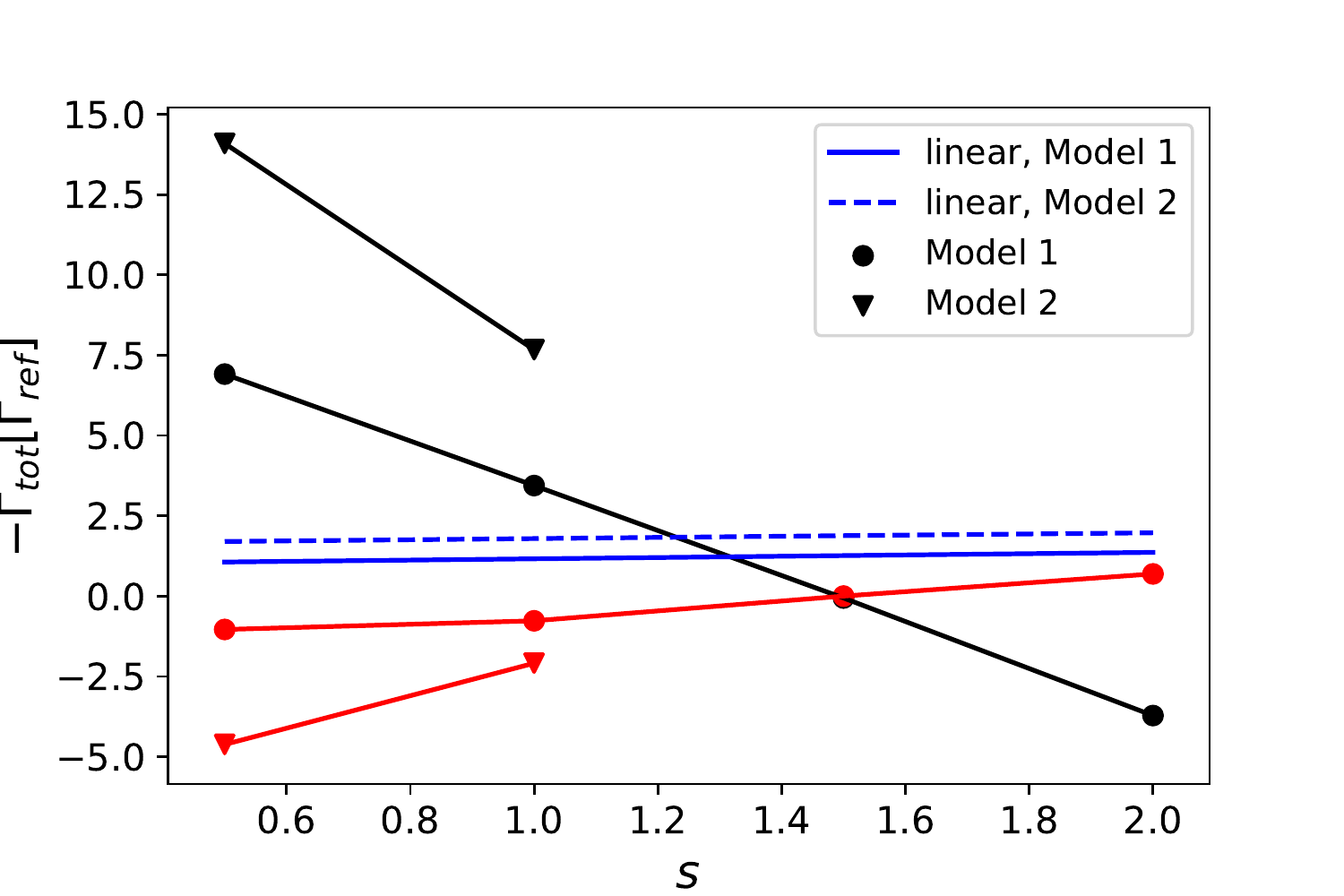}
\caption{Total Lindblad (black) and corotation (red) \textit{inward} torques for all our 2D cases obtained from summation of torques. Blue solid and dashed lines indicate linear total (corotation+Lindblad) torque obtained from applying Eqn \ref{Paardekoopertorque}.}
\label{fig:torquesummary}
\end{figure}

This large $h_p (=0.05)$ is comparable to $R_H/r_p = (q/3)^{1/3} \simeq 0.07$ for the $q=10^{-3}$ model parameter.  Consequently, the gap profiles of additional runs are much shallower than those for the smaller $h_p(=0.03)$ 
in Model 1 (Fig \ref{fig:model1}). The torques calculated from profiles are summarized in Fig \ref{fig:torquesummary}. Solid lines are planetary torques excited by the planet, and blue lines show predicted \textit{total}(corotation+Lindblad) linear torque from Eqn \ref{Paardekoopertorque} for the different smoothing lengths of Model 1 and Model 2, and extrapolated with $\Sigma_{min}/\Sigma_{p}$. {Red lines} are for corotation torques and black lines are for Lindblad torques. For gap-opening planets, they apparently deviate from the extrapolated scaling, and even outward migration is allowed in some cases. We note that these results themselves predict very well the direction and magnitude of the initial velocity at the instance of planet's release.

\section{Supplementary Simulation with \texttt{LA-COMPASS}}
\label{sec:LA-COMPASS}

We use an independent \texttt{LA-COMPASS} code to verify the above numerical results obtained with the
2D \texttt{FARGO} code.  We also use it to perform time dependent simulations as well as 3D steady
state simulations.

\subsection{2D \texttt{LA-COMPASS} simulation}
\label{sec:2dlacom}
\begin{figure*}[htp]
\centering
\includegraphics[width=0.48\textwidth]{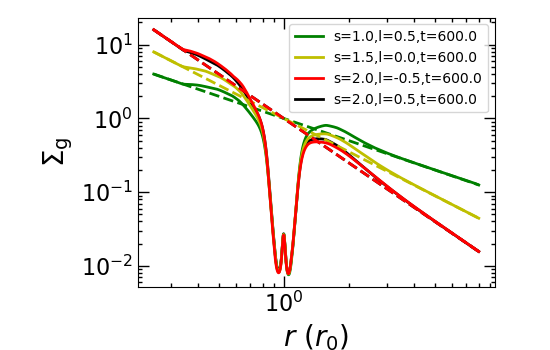}
\includegraphics[width=0.48\textwidth]{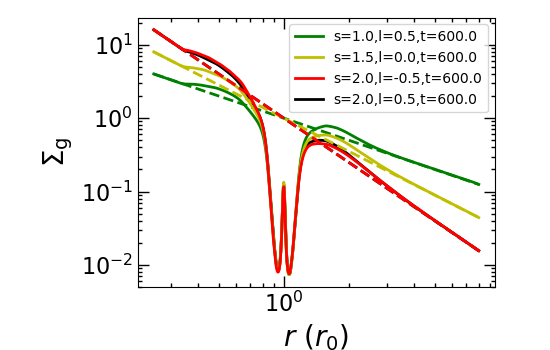}
\includegraphics[width=0.48\textwidth]{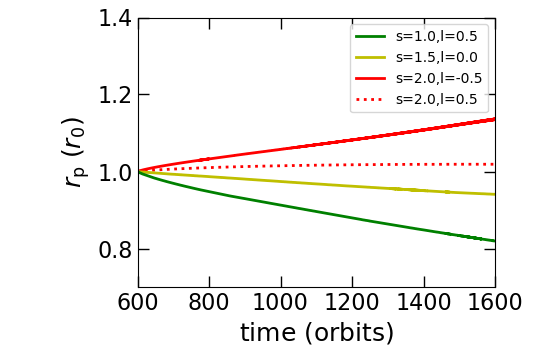}
\includegraphics[width=0.48\textwidth]{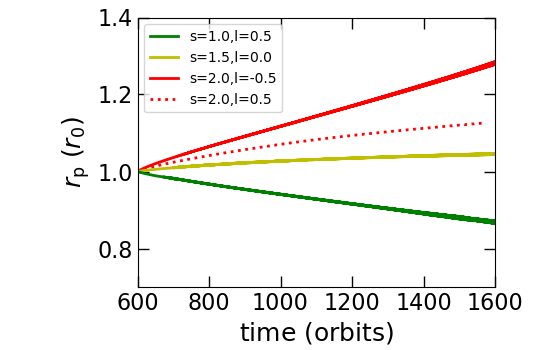}
\caption{Similar to Fig \ref{fig:model1}, but simulated with LA-COMPASS. For the extra case of $s=2.0$ and $l=0.5$ (black lines in upper panel and dotted lines in lower panel), we adopt a viscosity description of $\alpha=\alpha_{0}(r/r_{0})$, where $\alpha_{0}=1.11\times10^{-2}$. Left panels are for a resolution of $1024\times1256$, while the right panels are for $2048\times2048$.}
\label{fig:model1_lacompass}
\end{figure*}

In 2D simulation with \texttt{LA-COMPASS} \citep{Lietal2005,Lietal2009}, we use the same model parameters 
and boundary conditions as those in Model 1 using the \texttt{FARGO} code, for $s=1.0, s=1.5$ and $s=2.0$. 
The results are shown in Fig \ref{fig:model1_lacompass}.  The initial migration direction and pace for the standard resolution ($1024\times1256$) conform with 2D simulations of \texttt{FARGO}. Since the case with $l\leq 0$ may be unrealistic, we simulate another extra run for slope $s=2$, with $l=0.5$ and an increasing $\alpha \propto r$ to keep steady accretion (dotted lines). The very initial ($t=600$) profiles and the initial migration speed/direction in this case is similar to the constant $\alpha$ case for $s=2$, although change in $\alpha(r)$ give rise to a modest deviation as the location of planet shifts. Therefore, the torque components at planet release for $s=1.5,l=-0.5$ case in Fig \ref{case34} also applies to the $s=1.5,l=0.5$ more realistic case. We tested that for a higher resolution ($2048\times2048$), the total inward torques are slightly diminished, with $s=1.5$ case keeping a slow \textit{outward} migration rate from the very start. {This difference may arise from the torque cutoff at $0.6R_H$ being more precise for higher resolution \citep[see][Fig 1 for illustration]{DurmannKley2017}.} Since the predicted Lindblad \& corotation torque obtained from the gap profile for $s=1.5$ case is nearly zero (see Fig \ref{fig:torquesummary}), the actual migration direction is expected to be sensitive to small asymmetries in the CPD region.

Generally, the spikes in gas density profile inside the CPD region in \texttt{LA-COMPASS} are found to be less prominent than Fig \ref{fig:model1} and also dependent on resolution, {but this does not affect the migration torques in either numerical simulations or gap-profile calculations, since we did not include torques from this very small region in planet migration, and the initial gap profiles after CPD exlusion are very similar. }

{We have also verified the results for Model 2 in \S \ref{Model 2} with \texttt{LA-COMPASS}(not shown), and confirmed the gap profiles and migration rates for $s=0.5$ and $s=1.0$ are similar. \footnote{In an extra case with $s=1.5$ for Model 2, we find that the \texttt{LA-COMPASS} and \texttt{FARGO} results does not conform with each other well. The origin of this discrepancy is may arise from non-linear corotational torques or numerical issues, so we only showed the $s=0.5, 1.0$ cases for Model 2 in \S \ref{Model 2}, for which two codes yield similar results.}}

\begin{figure*}[htp]
\centering
\includegraphics[width=0.32\textwidth]{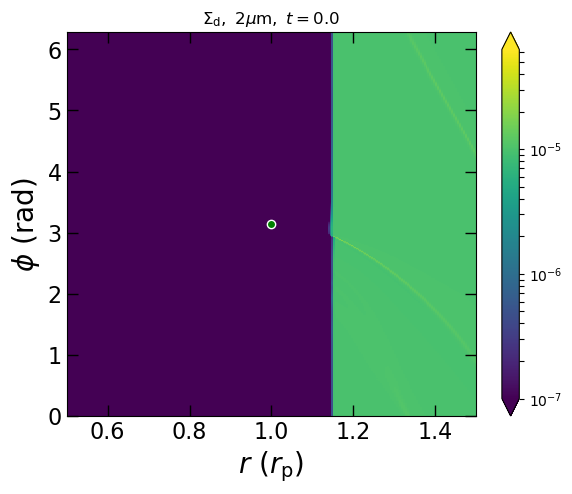}
\includegraphics[width=0.32\textwidth]{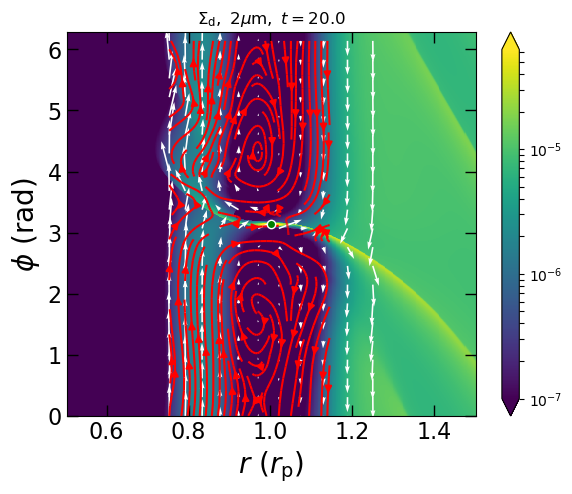}
\includegraphics[width=0.32\textwidth]{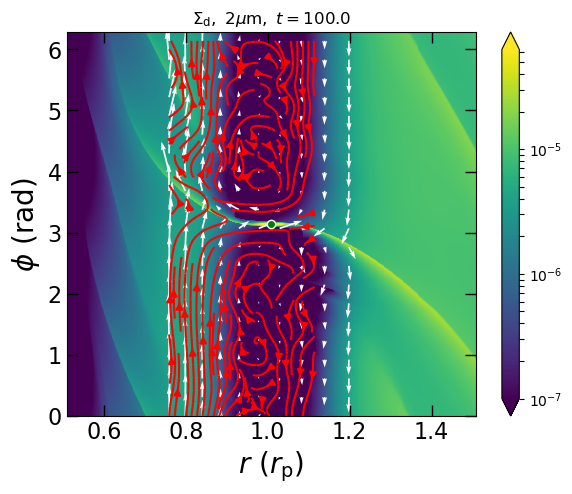}
\caption{Snapshots of tracer particles (simulated with well coupled dust particles with a small dust size of $2.0\ \mu{\rm m}$). The tracer particles are initially located outside 1.15 at 600 orbits (t=0 corresponds to the release time of 600 orbits.). White arrows indicate the velocity field of tracers, while red arrows show the streamlines of gas. Green dots are the location of the planet.}
\label{fig:tracer_lacompass0}
\end{figure*}

In the \texttt{LA-COMPASS} 2D simulation of Model 1 with $s=1.5$ we also examine the details of the gas flow across the disk by tracing density of heuristic particles initially distributed outside $r$=1.15, just beyond the outer edge of the gap. These particles are $\mu$m size and well coupled to the motion of the gas.  The evolution of particle surface density (analogous to Fig \ref{fig:tracers_fargo}) and the velocity field for these particles in the track structures (white arrows) are shown in Figure \ref{fig:tracer_lacompass0}. The red arrows show the streamline of the residual gas within the gap. 

While the residual gas originally trapped inside the horseshoe region continues to circulate around L4 and L5 points, the incoming gas/tracer particles simply flow across the gap driven by viscous diffusion bypassing the depleted region. They follow the contracting horseshoe orbits once they diffuses into the gap, and leave the horseshoe region on open streamlines as soon as they reach the \textit{inside} track of its horseshoe orbit which intersects with the inner edge (also see \S \ref{maintenance}). This effect is similar to that \citet{Lubowetal1999} found in the extreme case of a depleted inner disk (See their Fig 9\&10, where material cross the gap region and joins the inner flow in one horseshoe orbit). We attribute this streaming pattern to be the mechanism for gas to flow across the gap. Even from the inner gap edge to the mostly undisturbed region beyond it (interior to the gap), materials complete their diffusion process in $\sim 1$ horseshoe orbit.  Although a significant amount of gas on the horseshoe streamlines flow pass the planet's proximity, only a small fraction of it enter and stay within the planet's
Hill's sphere.  This flow pattern implies that accretion by the planet is inefficient (see further discussions 
in \S\ref{sec:massgrowth}).

\subsection{3D Simulation}

In 3D models, we only additionally specify an initial density structure in the vertical direction:
\begin{equation}
    \rho_{\rm g}(r,z)=\rho_{0}\left(r/r_{\rm p}\right)^{-\beta_{\rm 3D}}\exp\left({-z^2/2H^2}\right),
\end{equation}
where $\beta_{\rm 3D}=s+(3-l)/2$ and $\rho_{0}=\Sigma_{\rm p}/2\pi H(r_{\rm p})$ are chosen to match the 2D simulation parameters of Model 1 for $s=1,1.5,2$ cases with constant $\alpha$. We maintain the approximation that the disk is locally isothermal and temperature has no dependence on height.

\begin{figure}[htp]
\centering
\includegraphics[width=0.45\textwidth]{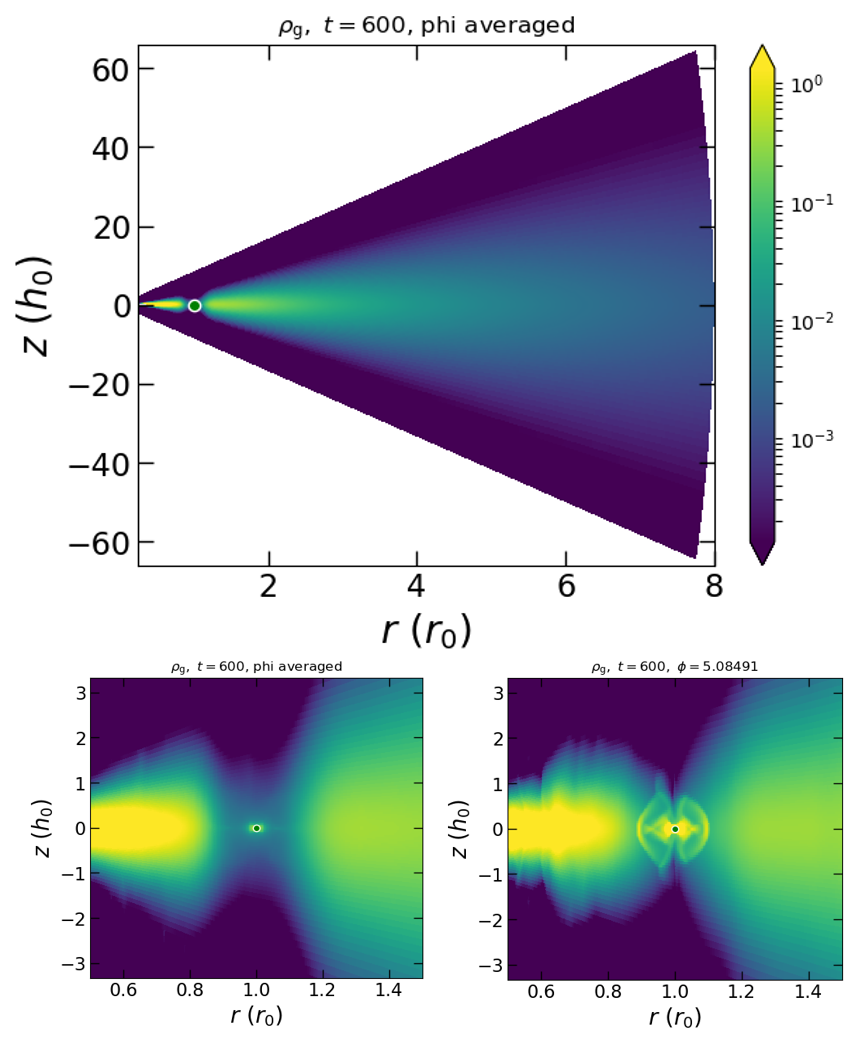}
\includegraphics[width=0.45\textwidth]{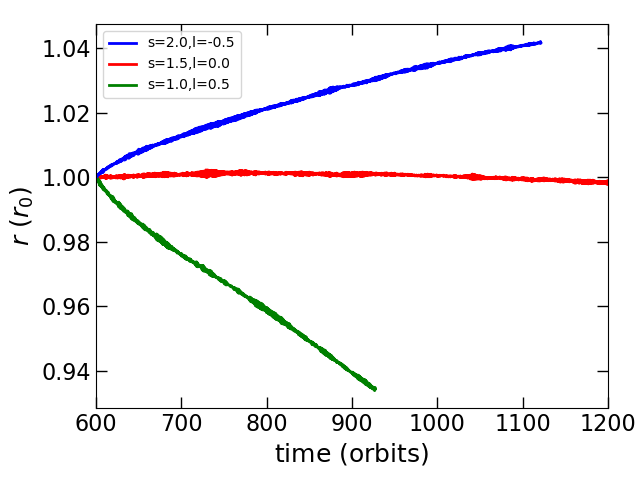}
\caption{Upper panel: vertical distribution of gas density at 600 orbits for 3D simulation with \texttt{LA-COMPASS}, corresponding to $s=1.5$ case of Model 1. Middle panel: zoom in vertical distribution of gas density around the planet orbital radius, left is azimuthally averaged values and right is the values at the exact planet azimuth. Lower panel: Planet orbital evolution of the Jovian planet in $s=1,1.5,2$ case of Model 1, in 3D simulations.}
\label{fig:3d_lacompass}
\end{figure}

In type I migration of super Earths, the linear torque from a 3D PPD can be different compared to 2D case \citep{Tanakaetal2002} due to resonances extending in the vertical direction. In contrast, the numerical simulations \citep{Fungetal2014} show that for the more massive gap-opening (Jovian mass) planets, the 3D and 2D gas profiles and planetary torques are very similar, at least when the 2D simulations are carried out with a softening length. In our 3D simulations (Fig \ref{fig:3d_lacompass}), we confirm that our results of initial migration speed and direction are also similar to those obtained with the 2D simulations.%, only that they diverge after a few hundred orbits. For example, in the $s=1.5$ case, the planet makes a transition from slowly outward migration to slowly inward migration after 200 orbits.

\section{Migration \& growth in evolving disks}
\label{sec:evolution}

In the previous sections, we adopt a steady state disk structure and found that the migration direction
is determined by the power index $s$ and pace is determined by $h_p$ and $R_H$. Negative torques acting on embedded 
gas giant planets tend to be reduced or reversed in regions where 
$s \geq 1.5$ (see \S\ref{sec:results}). Although this condition is difficult to realize in a steady state 
disk with a constant $\alpha$, such a 
steep fall off is possible in the outer regions of an evolving disk.  Here we construct an analytic 
disk evolution model and show that gas giants migrate outwards/inwards in the outer/inner regions 
of the disk. Such boundary conditions are physically more realistic.

\subsection{Self-similar disk evolution}
%For the steady state disk models, we used steady-state power-law disks for hydrodynamical simulations. Here we briefly discuss our results' implications on planet migration in evolving composite disks. 
In the absence of any planetary perturbation, the axisymmetric governing continuity and azimuthal momentum equations (Eqns \ref{eq:continuity} and \ref{eq:equationofmotion}) can be organized to be become
\begin{equation}
\frac{\partial \Sigma_{\mathrm{g}}}{\partial t}+\frac{1}{r} \frac{\partial}{\partial r}\left(\Sigma_{\mathrm{g}} v_{r} r\right)=0,
\label{eq:continuityeq}
\end{equation} 
\begin{equation}
    \Sigma_{\mathrm{g}} v_{r} r \frac{\partial\left(r^{2} \Omega\right)}{\partial r}=\frac{\partial}{\partial r}\left(r^{3} \Sigma_{\mathrm{g}} \nu \frac{\partial \Omega}{\partial r}\right).
\label{eq:eq_of_motion_phi}
\end{equation}
Under the approximation that viscosity is assumed to be a function of radius only (i.e. independent of 
$\Sigma$), they
reduce to a single linear diffusion equation with a set of analytic self-similar solutions
\citep{LydenBellPringle1974,Hartmannetal1998} such that

\begin{eqnarray}
 & {\displaystyle 
\Sigma_{\rm g} = \frac{M_{\rm d,0}}{2 \pi r_{\rm d}^2} 
\left(\frac{r}{r_{\rm d}}\right)^{-1} \tilde{t}^{-3/2}
                   \exp \left( -\frac{r}{\tilde{t} 
r_{\rm d}} \right),} \label{eq:self_similar3}\\
 & {\displaystyle \tilde{t} = \frac{t}{t_{\rm diff}}+1; 
\ \ \ \ \ {\rm and} \ \ \ \ \ 
t_{\rm diff}=\frac{r_{\rm d}^2}{3 \nu_{\rm d}}},
\label{eq:self_similar4}
\end{eqnarray}
where $M_{\rm d}(=\int_0 ^\infty 2 \pi r \Sigma_{\rm g} dr$) is the total disk 
mass and $M_{\rm d,0}$ is its initial ($t=0$) value.  Correspondingly, we have
\begin{equation}
v_r = - \frac{3\nu}{2r} \left(1- \frac{2r}{\tilde{t}r_{\rm d}}\right)
= - \frac{3\nu(r)}{2r} + \frac{3\nu(r_{\rm cr})}{2r_{\rm cr}}
\frac{r}{r_{\rm cr}},
\label{eq:vr_SS}
\end{equation}

\begin{equation}
\begin{aligned}
\dot{M}  =& - 3 \pi \Sigma_{\rm g} \nu \left(1- \frac{2r}
{\tilde{t}r_{\rm d}}\right) \\
=&
- {3 M_{d, 0} \nu_d \over 2 r_d^2} 
\left(1- \frac{2r}{\tilde{t}r_{\rm d}}\right) \tilde{t}^{-3/2} {\rm exp}
\left( - {r \over \tilde{t} r_d} \right),
\end{aligned}
\label{eq:mdotsim}
\end{equation}
where $r_d$ is the scaling radius of the disk.  
Disk gas flows inward at $r < r_{\rm cr} \equiv 
\tilde{t}r_{\rm d}/2$, but diffuses outward otherwise.
The critical radius $r_{\rm cr}$ expands with time,
especially at $t > t_{\rm diff}$.
Since $v_r \propto r$ for $r \gg r_{\rm cr}$,
$v_r$ diverges in the limit of $r \rightarrow \infty$.
But, because $\Sigma_g \propto \exp(-r/r_{\rm cr})$, $\dot{M}$
decays to zero in the limit of $r \gg r_{\rm cr}$.

\subsection{Irradiative outer disk}
\label{irradiative}
The simple-to-use self-similarity solution is directly applicable to the outer, irradiated 
region (B) of the disk where the gas temperature is determined by 
stellar irradiation. The temperature in this region can be approximated by the equilibrium
black body temperature
\begin{equation}
T_{\rm B} = \left( {L_\ast \over 4 \pi \sigma r^2} \right)^{1/4}
= T_{\rm AU} r_{AU}^{-1/2}
\label{eq:Tirra}
\end{equation}
where $r_{AU}\equiv r / 1 AU$, and $\sigma$ is the Boltzmann constant. The normalization temperature is 
\begin{equation}
T_{\rm AU} = \left( {L_\odot \over 4 \sigma AU^2} \right)^{1/4} \left(
{L_\ast \over L_\odot} \right)^{1/4} = T_\oplus l_\ast^{1/4},
\end{equation}
where $T_\oplus$ is the temperature at 1AU around the Sun, $l_\ast= L_\ast / L_\odot
\simeq m_\ast^2$ for T Tauri stars.  For the {\it ad hoc} $\alpha$ prescription, 
\begin{equation}
\nu = \alpha 
c_s^2/\Omega = \nu_{\rm d} (r/r_{\rm d}) 
\end{equation}
where
\begin{equation}
\begin{aligned}
    \nu_{\rm d} =& \alpha h_{\rm d} ^2 (G M r_{\rm d})^{1/2} \\
= &(G M_\odot {\rm AU})^{1/2} h_\oplus^2 \alpha r_{d, {\rm AU}} 
l_\ast^{1/4} m_\ast^{-1/2} , \\
h_{\rm d}^2 =&  (H/r)_{\rm d} ^2 = h_{\rm \oplus}^2 r_{d, {\rm AU}}^{1/2} 
l_\ast^{1/4} m_\ast^{-1},  
\end{aligned}
\end{equation}
$r_{d, {\rm AU}}= r_d/ 1 {\rm AU}$, and $h_{\rm \oplus}^2 = (R_g T_{\oplus}/ \mu) 
(1 {\rm AU} / G M_\odot)$.  This viscosity prescription is consistent
with the approximation which leads to the self similarity solution.

%(L_\ast/L_\odot)^{1/4} (M_\ast/M_

The surface density is determined from Eqn \ref{eq:self_similar3} such that 
\begin{equation}
\Sigma_{\rm B} = \Sigma_{\rm B0} 
%\left({M_\ast \over M_\odot}\right)^{-1/2} 
\left({r_{\rm AU}}\right)^{-1} \tilde{t}^{-3/2}
 \exp ( - r/\tilde{t} r_{\rm d} ), 
\label{eq:SigmaB}
\end{equation}
where 
\begin{equation}
\Sigma_{\rm B0} = \frac{M_{\rm d,0}}{2 \pi AU^2 r_{\rm d, AU}}
\end{equation}
    and
\begin{equation}
t_{\rm diff} =  {P_\oplus \over 6 \pi h_\oplus^2}
{r_{\rm d, AU} m_\ast ^{1/2} \over  \alpha  l_\ast^{1/4}}.
\end{equation}
In the outer irradiated region B, the slow-changing $r$-dependent surface density and temperature gradient is
\begin{equation}
s_{\rm B} \equiv - {\partial \ln \Sigma_{\rm B} \over \partial \ln r} =
1+\frac{r}{\tilde{t} r_{\rm d}},  \ \ \ \ \ \ 
l_{\rm B} \equiv - {\partial \ln T_{\rm B} \over \partial \ln r} 
={1 \over 2}
\end{equation}
as defined in Eqns \ref{eq:hr} and \ref{eq:sigmar}. Our 
2D \texttt{FARGO}, 2D and 3D \texttt{LA-COMPASS}
simulation indicate that the negative torques on embedded 
gas giant planets tend to be reduced or reversed in regions where 
$s_{\rm B} \geq 1.5$ (see \S\ref{sec:results}).  This condition is satisfied at $r \geq \tilde{t} 
r_{\rm d}/2= r_{cr}$.  Thus, the migration direction of an embedded gas giant
is consistent, by coincidence in this particular model,  with that of viscous diffusion of the disk gas.

\subsection{Inner regions heated by viscous dissipation}

The inner region A is heated by viscous dissipation and should have a different structure compared to region B \citep{GaraudLin2007}.  \citet{Ali-Dibetal2020} developed simple analytical models of steady-state active disks with power-law $T$ and $\Sigma$ profiles. Although an approximately constant $\Sigma v_r r$ radial distribution, is established, the structure of the inner region A evolves during the depletion and spread of the disk.
Here we generalize their ``radiative disk" model for a quasi steady self-similar profile. The equation of motion (Eq \ref{eq:eq_of_motion_phi}) implies
\begin{equation}
\left( \Sigma \nu \right)_{\rm A} 
= - {r^{-1/2} \over 3} \int \Sigma_{\rm A} v_r r^{-1/2} dr.
\end{equation}

According to the self consistent $\alpha$ prescription, the effective 
viscosity is no longer proportional to $r$ only (it also depends on 
$\Sigma$) such that the \textit{classical} self similarity 
solution is not strictly applicable.  However, the distribution of $\dot M$ 
in most regions of the disk can still be adequately approximated by
\begin{equation}
\left( \Sigma \nu \right)_{\rm A} = \left( \Sigma \nu \right)_{\rm A0}
\tilde{t}^{-3/2} \exp \left( - \frac{r}{\tilde{t} r_{\rm d}} \right)
\end{equation}
where
\begin{equation}
\left( \Sigma \nu \right)_{\rm A0}
= {\alpha h_\oplus^2 \over P_\oplus} 
{M_{\rm d,0} \over m_\ast^{1/2}}  {l_\ast^{1/4} \over r_{\rm d, AU}}.
\end{equation}
Assuming that the energy in the inner region is generated by viscous heating and then vertically transported by radiative diffusion, the energy equation reduces to

\begin{equation}
{9 \over 4} \left( \Sigma \nu \right)_{\rm A} \Omega^2 
= {2 \sigma T_{\rm A}^4 \over \kappa_0 T_A (\Sigma_A/2)}
\label{eq:evis}
\end{equation}
where $\kappa = \kappa_0 T$ is an approximation for the 
grain opacity, neglecting sublimation \citep{GaraudLin2007}.  With the $\alpha$ prescription for
viscosity ($\nu=\alpha h^2 
\Omega r^2=\alpha R_{g} T / \Omega$ where $R_{g}$ is the gas constant), we find
\begin{equation}
\begin{aligned}
T_{\rm A} &= \left[ {9 \kappa_0 \Omega^3 \over 16 \sigma
\alpha R_g} \left( \Sigma \nu \right)_A^2 \right]^{1/4}\\
&= T_{\rm A0} r_{AU}^{-9/8}
\tilde{t}^{-3/4} \exp \left( - \frac{r}{2\tilde{t} r_{\rm d}} \right)
\end{aligned}
\end{equation}

\begin{equation}
\begin{aligned}
T_{\rm A0}=& T_\oplus \left( {3 h_\oplus \over 2} \right)^{1/2}
\left[ {M_{d,0} V_\oplus^2 \over L_\odot 
P_\oplus} {\alpha \kappa_0 T_{\oplus} M_{d, 0}
\over 2 AU^2 } \right]^{1/4} \\ \times&
\left( {m_\ast^{1/8} l_\ast^{1/8}\over r_{d, AU}^{1/2} } \right) 
\end{aligned}
\end{equation}
where the normalization speed and time, $V_\oplus={\sqrt{GM_\odot/1 {\rm AU}}}$ and 
$P_\oplus= 2 \pi {\rm AU} / V_\oplus$ are the Keplerian velocity and period of the 
Earth respectively.  We also find 
\begin{equation}
\Sigma_{\rm A} = { (\Sigma \nu)_{\rm A} \Omega \over \alpha R_g T_{\rm A}}
= \Sigma_{\rm A0} r_{AU}^{-3/8}
\tilde{t}^{-3/4} \exp \left( - \frac{r}{2\tilde{t} r_{\rm d}} \right)
\end{equation}
 \begin{equation}
 \begin{aligned}
\Sigma_{\rm A0}&= {M_{d, 0} \over 2 \pi AU^2 r_{d, AU}}
\left( {2 \over 3 h_\oplus} \right)^{1/2}\\
&\times
{1 \over m_\ast^{1/8} r_d^{1/2} }
\left[ {M_{d,0} V_\oplus^2 \over L_\odot 
P_\oplus} {\alpha \kappa_0 T_{\oplus} M_{d, 0}
\over 2 AU^2 } \right]^{-1/4} l_\ast^{1/8}.
\end{aligned}
\end{equation}
In the inner region A, 
\begin{equation}
\begin{aligned}
s_{\rm A} \equiv -{\partial \ln \Sigma_{\rm A} \over \partial \ln r} =&
\left({3 \over 8}+\frac{r}{2 \tilde{t} r_{\rm d}} \right)\\
l_{\rm A} \equiv -{\partial \ln T_{\rm A} \over \partial \ln r} 
=&\left( {9 \over 8} + \frac{r}{2 \tilde{t} r_{\rm d}} \right).
\label{eq:sA}
\end{aligned}
\end{equation}
The significant reduction or reversal of inward migration occurs (when $s_{\rm A}\geq1.5$) in region A at 
$r\geq 9 \tilde{t} r_{\rm d}/4 =4.5 r_{\rm cr}$.

Transition radius $r_t$ between these two regions (A and B) occurs at a radius where $T_{\rm A}
= T_{\rm B}$ such that
\begin{equation}
r_{t, AU} ^{-5/8} \exp \left( - {r_t \over 2 \tilde{t} r_{\rm d}} \right)
= {T_{\rm B 0} \over T_{\rm A0}} \tilde{t}^{3 /4} \left( {M_\ast \over
M_\odot} \right)^{1/8}.
\end{equation}
As the disk gas depletes with time, the viscously heated region shrinks and $r_t$ decreases
with time.  Although $r_t$ may be initially comparable to $r_{\rm cr}$, it is unlikely for
it to exceed $4.5 r_{\rm cr}$. Thus, gas giants with $R_H \gtrsim H$ general migrate inwards
in region A.  As the irradiated region B enlarges with time, the determination factor for the
migration direction becomes $s_{\rm B}$.  Gas giants migrate outward/inward in regions 
exterior/interior to $r_{\rm cr}$. 

\subsection{Numerical simulation of planets in depleting disks}

\begin{figure}[htp]
\centering
\includegraphics[width=0.48\textwidth]{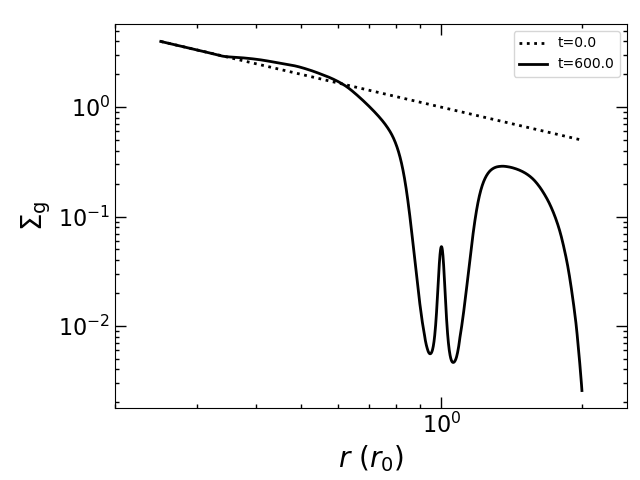}
\includegraphics[width=0.48\textwidth]{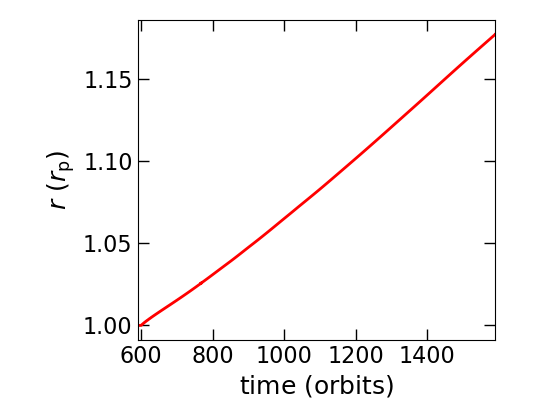}
\caption{We initially set $s=1.0$ and $l=0.5$ similar to the case of Model 1, but with a very small outer boundary $r_{\rm out}=2.0$ of the disk. We only allow outflow from the outer boundary, while the inflow from the outer boundary is disabled and damping condition is still imposed.  The upper panel shows the azimuthal averaged gas surface density at $t=0$ (dotted line) and $t=600$ (solid line). The lower panel shows the orbital evolution of the planet. }
\label{fig:bc_lacompass}
\end{figure}

In Fig \ref{fig:bc_lacompass},  we present results of {additional \texttt{LA-COMPASS} numerical simulation} of a Jupiter-mass planet
(with $q=10^{-3}$) in case 2 with the parameters of Model 1. Top panel shows the unperturbed surface 
density and the gap profile at the instant of planet's release.  The lower panel shows the orbital 
evolution. In contrast to all the above models, only outflow is allowed from the outer boundary and 
the inflow is disabled in this case.  The objective of this model is to simulate the depletion 
of the disk gas analogous to that approximated by the self-similar outer-disk solution in the previous
subsection. We set $r_{\rm out}=2.0$ to let the depletion quickly reach the planet's orbital radius.
All the OLRs are still included in the simulation domain. For $s=1.0$ and $l=0.5$ of steady state Model 1,
the planet is originally migrating inwards rapidly (Fig \ref{fig:model1}), however in a depleting disk, 
the $\Sigma$ distribution quickly changes. Outside the planet's orbit, the $\Sigma$ gradient become steeper and the planet migrates outwards after $t=600$ orbital period.

\subsection{Implication on the planet's final mass and location}
\label{sec:massgrowth}
The condition for steady state flow across the gap might break down if the embedded planet continues
to gain mass and attains a $q>q_{ss}$ (Eqn \ref{eq:qss}).  Despite the presence of gap-crossing streaming motion, 
gas accretion onto embedded planet is possible interior to both its Bondi radius ($R_B = G M_p/2 c_s^2(r_p)$)
and Hill's radii ($R_H$) even though we neglected this effect in the hydrodynamical simulations. In reality, the planet gas accretion is limited by Kelvin-Helmholtz (KH) contraction for small $q$ until runaway gas accretion shortens the KH timescale, and the gas accretion becomes hydrodynamically limited \citep[e.g.][]{PisoYoudin2014,Leeetal2014,LC2015,GinzburgChiang2019,Ali-Dibetal2020,2020ApJ...896..135C}. At the critical asymptotic mass $q_{am} \sim 2 {\sqrt {2/3}} h^3$, $R_B \simeq R_H$.
This mass is comparable to the thermal mass $q_{th} \sim 3 h^3$ for which $R_H \sim H$ and above which
the protoplanets’ tidal perturbation induces the formation of a gap \citep[e.g.][]{Papaloizou_Lin_1984, Ward_1997}. For $q< q_{am}\sim q_{th}$, $H>R_H>R_B$ and the sub-thermal planet accretes at the Bondi rate \citep{Franketal1992}:

\begin{equation}
    \dot M_{Bondi} \sim \dfrac{\Sigma_{min}}{H} c_s(r_p) R_B^2
\end{equation}

For $q> q_{th}\sim q_{am}$ super-thermal planets which we focus on in this paper, the hierarchy becomes $R_B>R_H>H$ and usually a Hill radius-determined accretion rate is imposed to be 

\begin{equation}
    \dot{M}_{ Hill} \sim \Sigma_{min} R_{H}^2 \Omega 
\end{equation}

The accretion rates depends on the surface density in the gap region $\Sigma_{min}$, which is always maintained for planets just grown to open a gap, since $q_{ss}$ is an order of magnitude larger than $q_{am}$, and steady-state inwardly diffusing mass flux can be maintained across the gap. 

Under the assumption that gas flow across the gap and the planetary accretion is maintained until disk depletion, and neglecting 2D/3D asymmetries, \citet{Rosenthal2020} estimate, based on the equivalent minimum density scaling in Eqn \ref{bottomdensity}, that a gap-opening planet (in a normal viscous environment) can accrete surrounding materials very efficiently, until it reaches a critical mass (see their Eqn 31) when the accretion rate is finally reduced:

\begin{equation}
    q \geq q_{r} = 3.6 h_p^{9/4}
\end{equation}

This critical mass is much larger than the thermal mass, which implies that gap-opening gas giants can still efficiently consume a considerable amount of disk mass flow as $q_{th}<q<q_r$. \citet{TanigawaTanaka2016} also estimated, based on the empirical scaling of \citet{TanigawaWatanabe2002}, that most of the global gas flow might be accreted by a Jupiter-mass planet in a gas-poor disk. For such accretion rates, the gas profile would be moulded by planet accretion and its migration would also be affected.

However, 2D hydrodynamical simulations show that the net azimuthal mass flux into the protoplanet’s Hill radius 
is reduced after $q \geq q_{am}$ \citep[][see their Eqn 28 for the exponential reduction]{Dobbsdixon2007} despite the presence of gas along the horseshoe streamlines in the gap. The main physical reasons for this transition to inefficient planetary accretion are: 
1) the fraction of gas on the horseshoe streamlines which can reach, in azimuth, the planet's Hills sphere 
before making a U-turn decreases with the planet's $q$. 2) Moreover, the mismatches in both the vortensity \citep{Papaloizou1989} and Bernoulli energy in the circumstellar horseshoe and circumplanetary disk flows
increases with the planet's $q$.  3) While both vortensity and Bernoulli energy are conserved along the streamlines \citep{Korycansky1996, Balmforth2001}, flow from the horseshoe streamlines into the protoplanet's Hill's sphere and the gas accretion onto the circumplanetary disks requires shock dissipation.  Consequently, only a fraction of the gas pass through this region is retained \citep{Dobbsdixon2007}.  These previous results 
reconfirmed by the results in Figure \ref{fig:tracer_lacompass0} which shows that only a fraction of the 
tracer particles is actually captured into and retained by the planet's Hills radius. In a subsequent paper, we will use numerical simulations to investigate the properties of this gas giant accretion barrier in more details.

%also investigated gas accretion and final masses of gas giants, using empirical accretion rate formula from \citep{TanigawaWatanabe2002} combined with a gas density formula similar to Eqn \ref{bottomdensity} to find a large estimate of planet's final mass. In these papers, the estimated accretion rates for Jupiter might be very high

If the gas accretion rate onto the planet is severely quenched after it reaches thermal mass, its 
mass increase would be stalled with a final mass ratio $q_{am} < q_{final} < q_{ss}$. For modest 
value of $h$, both $q_{am}$ and $q_r$ are smaller than $q_{ss}$. For these asymptotic values, steady 
flow across the gap preserves the $\Sigma$ distribution for the low order Lindblad resonances outside the gap.  The planet's subsequent migration direction and pace are then determined by the surface density distribution in the unperturbed regions of the disk ($\Sigma$ and $s$). 

\subsection{Classical type II migration of massive planets}
\label{sec:classicaltype2}

\begin{figure}[htp]
\centering
\includegraphics[width=0.48\textwidth]{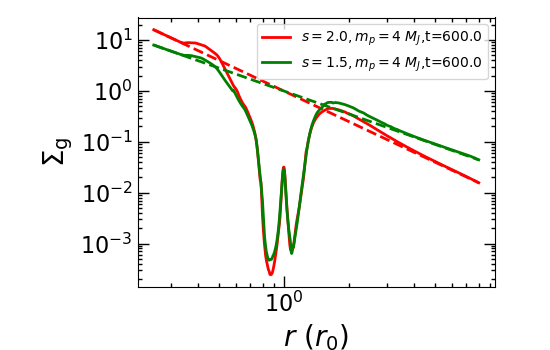}
\includegraphics[width=0.48\textwidth]{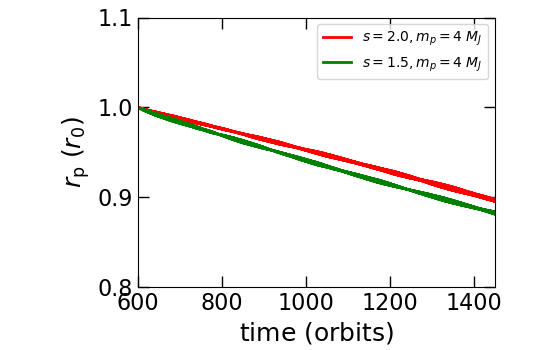}
\caption{Similar to the case of $s=1.5$, $l=0.0$ (green lines) and $s=2.0$, $l=-0.5$ (red lines) of model 1, but with a planet mass of $4\ M_{\rm J}$. The upper panel shows the azimuthal averaged gas surface density at t=0 (dotted line) and t=600 (solid line). The lower panel shows the orbital evolution of the planet. }
\label{fig:type2mig}
\end{figure}

Many gas giants are found in multiple systems.  These system may undergo dynamical instability which
leads to eccentricity excitation and mergers \citep{Lin1997, Ida2013}.  These processes can even lead to 
masses larger than $q_{ss}$.  Massive planets (with $q \gtrsim q_{ss}$) 
open deep gaps \citep{Bryden1999} which might interrupt the steady state flow. On the global viscous evolution time scale, 
disk gas interior to these planets is depleted whereas outside their orbit accumulates.  This evolution 
reduces/enhances the inner/outer Lindblad torque.  This torque imbalance drives the planet to migrate with the 
viscous evolution of the disk as in the classical type II migration \citep{Lin_Papaloizou_1986}, with direction determined by their radial position relative to the radius of maximum viscous stress $r_{\rm cr}$. This 
condition is similar to the lower-mass case (with $q_{final} < q_{ss}$) in outer regions of the disk (\S \ref{irradiative}), albeit the gas flow is cut off especially when the mass of the planet becomes comparable to or larger than the mass 
of the residual gas in the disk \citep{Lin_Papaloizou_1986}.

%\subsection{Numerical simulation of classical type II}

In Fig \ref{fig:type2mig} we present \texttt{LA-COMPASS} results of numerical simulation for $q>q_{ss}$ planets' migration. Red line is for a $q=4\times 10^{-3}$ planet in the $s=2.0, l=-0.5$ parameters of Model 1, while the green line is for $s=1.5, l=0$. The top panel shows the unperturbed surface densities (dashed lines) and the gap profiles (solid lines) at planet release, while lower panel shows the orbital evolution. In the cases of Jovian planets in these parameters, the planet is originally migrating very slowly inwards or outwards. However for 4 $M_{\rm J}$ cases, the planets are both migrating inwards with speeds similar to viscous velocity $\sim \nu (r_p)/r_p$ (see Eqn \ref{u_pvis}) as qualitatively inferred above. Also, in these $q>q_{ss}$ cases, we can clearly see signs of disk depletion on the inner side of the gap, as a result of gas flow across the gap being quenched.
%predicted in \S \ref{sec:classicaltype2}.

\section{Discussion}
\label{sec:summary}
\subsection{Pileups}
%If there is no gas flow across the disk, for inward p$u_{p,vis}\approx 7\times 10^{-5} \dfrac{\Omega_pr_p}{2\pi}$lanet migration faster than the accretion viscosity, a gas pileup in the inner disk would continue to grow while materials would be depleted outside the planet's orbit because the gas could not follow the planet's movement. However, such pileups are not expected to occur if gas can efficiently transport through the gap and the gap profiles are maintained.
\citet{Dempseyetal2020} proposes that for inward migrations where gas flow through the gap is very efficient and a steady gap profile is maintained, a total negative torque must be accompanied by a significant pileup in the \textit{outer} disk. In these states, a more accurate description of accretion rate than Eqn \ref{eqn:accretion rate} outside the wave deposition zone should be:

\begin{equation}
    \dot{M}+\dfrac{\Gamma_{dep}}{l}=3\pi \nu\Sigma
\end{equation}
where $l$ is the specific angular momentum of the disk and $\Gamma_{dep}$ is the total amount of torque deposited onto the flow, which equals to the \textit{total amount} of negative torque acting on the planet. Including such a description at the outer boundary might give a more self-consistent rate for rapid inward migration that differs on order of
\begin{equation}
   \left.\dfrac{\Gamma_{ref}}{\dot{M}l}\right|_{r_{max}} \sim \left.\dfrac{25 \Gamma_0}{K\dot{M}l}\right|_{r_{max}}\sim  25h_p\sqrt{\dfrac{r}{r_{max}}},
\end{equation}
from what we obtain from the classical boundary condition.  But this effect does not change the direction to outward migrations and/or our findings of delicate Lindblad resonance balance across a steep gap structure. Indeed pileup structures are noticeable in the results of our simulations, and we have already shown that continuous gas flow can maintain our gap profiles for a relatively long time (see \S \ref{maintenance}). When $\Gamma_{dep} \lesssim 0$, we expect a deficit in surface density of the outer disk as \citet{Dempseyetal2020}'s boundary conditions describe, which in turn boosts the outward migration. 

%Nevertheless, it's still important to confirm that we have efficient gas flow inwards across the gap for our cases where Lindblad torques are suppressed and the planet moves against viscous evolution. 

\subsection{Rayleigh instability}
For very sharp gap edges, the $\Sigma$ and $\Omega$ profiles might be significantly perturbed that the specific angular momentum marginally decreases outward \citep{Chandrasekhar1960}

\begin{equation}
    \dfrac{\partial l}{\partial r}=2rB<0.
\end{equation}

In these regions, Rayleigh instability induces turbulent gap streamers that tend to smooth out the profile \citep{Papaloizou_Lin_1984, deValborroetal2006}. But this effect only reduce the $\Sigma$ gradient to a certain extent \citep{Fungetal2014} and it does not eliminate the unstable region entirely, as suggested by some analytical gap models \citep{TanigawaIkoma2007,Kanagawaetal2015MNRAS}. In these regions, we expect there to be no real solution of epicycle frequency $\kappa=\sqrt{4\Omega B}$, which bring in further complexities to the resonances. However, we did not discover such regions for $q=10^{-3}$ and all $\kappa'$ are real for our gap profiles that go into torque calculations in \S \ref{sec:results}. Rayleigh-unstable regions may appear for planets with masses of a few Jupiters, the effects of which may be addressed by future research.

\subsection{Summary}
We have investigated the migration of Jovian planets ($q=10^{-3}$) due to disk-planet interaction in locally isothermal disks with 2D and 3D hydrodynamical simulations. The typical disk mass $4\pi \Sigma_p R_p^2$ in our simulations are larger than the planet mass so we do not go into the regime of type III migration \citep{PapaloizouMasset2003}. We neglect gas accretion onto the planet and suppress the torque from the circumplanetary disk which may lead to spurious runaway migration. 

Our main methods and results are presented respectively in \S \ref{sec:methods} and \S \ref{sec:results}. We analyzed the torque components and find that the total torque lies delicately on the balance between low-order Lindblad torques arising from the gap edges. In regions where gas profile is relatively flat, the $m$-th order ILR is located farther from the planet than the corresponding OLR in second order of $1/m$.  This difference results in the negative Lindblad torque dominating over the positive Lindblad torque in type I migration. However, in steep regions at the edges of a deep gap (large $K$), higher  surface  density  at distance farther from the planet makes up for this loss, and may let low order ILR dominate over OLR (Fig \ref{case34}). In cases of large $s$, this can lead to significant suppression of migration rate or even reversal of the migration direction, and long-period gas giants may be retained. This effect must be taken into consideration in future attempts to construct analytical gas profiles of gap-opening planets, for which the assumption that the bulk of deposited torques arise from the bottom of the gap \citep{Kanagawaetal2015MNRAS,Duffell2015b} breaks down, and we expect order-unity deviations from the bottom density formula. For shallower gaps with less steep density gradient at the gap edges (smaller $K$ and $s$), the bulk of the resonance arise from the bottom of the gap, and the total migration torque becomes closer to the simple scaling of extrapolating type I torque with $\Sigma_p \rightarrow \Sigma_{min}$ \citep{Kanagawaetal2018} and the depletion factor $\Sigma_{min}/\Sigma_p$ itself would be consistently given by $1/(1+0.04K)$.

In the analysis of torque components, we adopt the gas profile at the beginning of planet migration. Since the total torque is quite dependent on the shape of such profile, the planet has to maintain an initial profile in its vicinity as it migrates in order to have a stable migration rate. This structure is sustained by the gas flow across the gap.  We show in \S \ref{maintenance} (also see Fig \ref{fig:tracer_lacompass0}) that this can be done efficiently through loading materials at the \textit{outer edge} of the gap on onto horseshoe orbits, and diffusing it onto the inner disk when its contracting orbit (due to viscosity) grazes the \textit{inner edge} of the gap. The critical mass above which the steady flow could be maintained is given by $q_{ss} \approx 12.5 h^{18/7}$ for the approximate bottom density scaling of Eqn \ref{bottomdensity}. This critical mass estimate is applicable to order unity since we also expect deviations from the bottom density formula for gap-opening planets.

Apart from the main 2D simulations on \texttt{FARGO}, we have also carried out 2D and 3D simulations on \texttt{LA-COMPASS} to obtain similar results as verification in \S \ref{sec:LA-COMPASS}. In \S \ref{sec:evolution}, we applied self-similar analytical models and numerical simulations to discuss the migration of gas giants in evolving disks, the implications on final mass and location of giant planets. Despite the constant flow across the disk, an individual gas giant's accretion is quenched after reaching the thermal mass \citep{Dobbsdixon2007}.  The planet should attain a final mass of around $q_{th}<q_{final}<q_{ss}$ before the gas supply into the its Hill radius is severely quenched. Gas giants may also acquire $q>q_{ss}$ as a result of merger events, and their orbit may subsequently evolve in the direction and at the rate predicted by classical type II migration.

Our linear torque analysis for a given $\Sigma$, $\Omega$ profile is based on a 2D approximation \citep{GT1980,Ward1989,1993ApJ...419..155A}. Detailed calculation of the total torque in the 3D limit, 
based on the full gas density distribution obtained from 3D numerical simulations, as well as the
location and torque intensity of the Lindblad and corotation resonances\citep{Tanakaetal2002} are 
left for future research. {The detailed dependence of simulation results on numerical resolution also remains to be further investigated.}

\acknowledgements
We thank Shigeru Ida, Eugene Chiang, Adam Dempsey for useful discussions. We thank an anonymous referee who provided encouraging feedback. Y.X.C thanks the Tsinghua Spark Undergraduate Research program for support of this work and Yikai Wang for helpful assistance. 
Y.P.L. and H.L. gratefully acknowledge the support by LANL/LDRD. 
This research used resources provided by the Los Alamos National Laboratory Institutional Computing Program, which is supported by the U.S. Department of Energy National Nuclear Security Administration under Contract No. 89233218CNA000001. 
\textit{Softwares}: \texttt{FARGO} \citep{Masset2000}, \texttt{LA-COMPASS} \citep{Lietal2005, Lietal2009}, \texttt{Numpy} \citep{2011CSE....13b..22V}, \texttt{Scipy} \citep{2019arXiv190710121V}, \texttt{Matplotlib} \citep{2007CSE.....9...90H}
%\software

\bibliography{new}
\bibliographystyle{aasjournal}

\end{document}